\documentclass[aps,prd,10pt,twocolumn,superscriptaddress,showpacs,longbibliography]{revtex4-1}

\pdfoutput=1
\usepackage{graphicx}
\usepackage{amsfonts,amsmath,amssymb,bm,bbm}
\usepackage{color}
\usepackage{enumitem}
\usepackage{url}
\usepackage{multirow}
\usepackage{hhline}
\usepackage{soul}

\usepackage{hyperref}
\hypersetup{
    pdfnewwindow=true,
    colorlinks=true, 
    linkcolor=blue, 
    citecolor=blue,
    filecolor=blue,
    urlcolor=blue
}

\def\bi{\begin{itemize}[noitemsep,leftmargin=*]
\setlength\itemsep{1em}
        }
\def\ei{\end{itemize}}

\begin{document}

\title{Dimuons in Neutrino Telescopes: New Predictions and First Search in IceCube}  

\author{Bei Zhou}
\email{beizhou@jhu.edu}
\thanks{\scriptsize \!\!  \href{https://orcid.org/0000-0003-1600-8835}{orcid.org/0000-0003-1600-8835}}
\affiliation{William H. Miller III Department of Physics and Astronomy, Johns Hopkins University, Baltimore, Maryland 21218, USA}

\author{John F. Beacom}
\email{beacom.7@osu.edu}
\thanks{\scriptsize \!\!  \href{http://orcid.org/0000-0002-0005-2631}{orcid.org/0000-0002-0005-2631}}
\affiliation{Center for Cosmology and AstroParticle Physics (CCAPP), Ohio State University, Columbus, Ohio 43210, USA}
\affiliation{Department of Physics, Ohio State University, Columbus, Ohio 43210, USA}
\affiliation{Department of Astronomy, Ohio State University, Columbus, Ohio 43210, USA}

\date{arXiv v1: 6 October 2021; arXiv v2 and journal: 28 December 2021}

\begin{abstract}
Neutrino telescopes allow powerful probes of high-energy astrophysics and particle physics.  Their power is increased when they can isolate different event classes, e.g., by flavor, though that is not the only possibility.  Here we focus on a new event class for neutrino telescopes: dimuons, two energetic muons from one neutrino interaction.  We make new theoretical and observational contributions.  For the theoretical part, we calculate dimuon-production cross sections and detection prospects via deep-inelastic scattering (DIS; where we greatly improve upon prior work) and $W$-boson production (WBP; where we present first results).  We show that IceCube should have $\simeq 130$ dimuons ($\simeq 6$ from WBP) in its current data and that IceCube-Gen2, with a higher threshold but a larger exposure, could detect $\simeq 620$ dimuons ($\simeq 30$ from WBP) in 10 years.  These dimuons are almost all produced by atmospheric neutrinos.  For the observational part, we perform a simple but conservative analysis of IceCube public data, finding 19 candidate dimuon events.  Subsequent to our paper appearing, visual inspection of these events by the IceCube Collaboration revealed that they are not real dimuons, but instead arise from an internal reconstruction error that identifies some single muons crossing the dust layer as two separate muons.  To help IceCube and the broader community with future dimuon searches, we include the updated full details of our analysis. Together, these theoretical and observational contributions help open a valuable new direction for neutrino telescopes, one especially important for probing high-energy QCD and new physics.
\end{abstract}

\maketitle


\section{Introduction}
\label{sec_intro}

Neutrino observatories have been making groundbreaking discoveries~\cite{Davis:1968cp, Kamiokande-II:1987idp, Bionta:1987qt, Super-Kamiokande:1998kpq, SNO:2002tuh, Aartsen:2013jdh}.  For astrophysics, high-energy neutrino telescopes in particular can reveal the elusive sites of hadronic cosmic-ray acceleration.  IceCube has discovered TeV--PeV diffuse astrophysical neutrinos~\cite{Aartsen:2013jdh, Aartsen:2015knd, Aartsen:2020aqd, IceCube:2020wum} and a few promising sources~\cite{IceCube:2018cha, IceCube:2018dnn, Aartsen:2019fau}.  For particle physics, high-energy neutrino telescopes reach unprecedented energy scales, detecting some standard-model processes for the first time~\cite{Glashow:1960zz, Lee:1960qv, Lee:1961jj, Seckel:1997kk, Alikhanov:2015kla, IceCube:2017roe, Bustamante:2017xuy, Zhou:2019vxt, Zhou:2019frk, IceCube:2020rnc, IceCube:2021rpz, Robertson:2021yfz}.  They also provide unique information on new physics, including dark matter~\cite{Beacom:2006tt, Yuksel:2007ac, Murase:2012xs, Feldstein:2013kka, Esmaili:2013gha, Murase:2015gea, IceCube:2016dgk, DiBari:2016guw, IceCube:2018tkk, Arguelles:2019ouk, ANTARES:2020leh}, novel neutrino interactions ~\cite{Lipari:2001ds, Cornet:2001gy, Beacom:2002vi, Ng:2014pca, Ioka:2014kca, Shoemaker:2015qul, Bustamante:2016ciw, Denton:2018aml, Bustamante:2020mep, Creque-Sarbinowski:2020qhz, Esteban:2021tub}, and other exotic signals~\cite{Albuquerque:2003mi, Albuquerque:2006am, Ando:2007ds, ICECUBE:2013jjy, Kopper:2015rrp, Kopper:2016hhb, Meighen-Berger:2020eun, IC_stau_2021, Coloma:2017ppo}.  Much more progress is expected with new telescopes, including KM3NeT~\cite{Adrian-Martinez:2016fdl}, Baikal-GVD~\cite{Baikal-GVD:2018isr}, P-ONE~\cite{P-ONE:2020ljt}, and especially IceCube-Gen2~\cite{IceCube-Gen2:2020qha}.

\begin{figure}[t!]
\includegraphics[width=0.9\columnwidth]{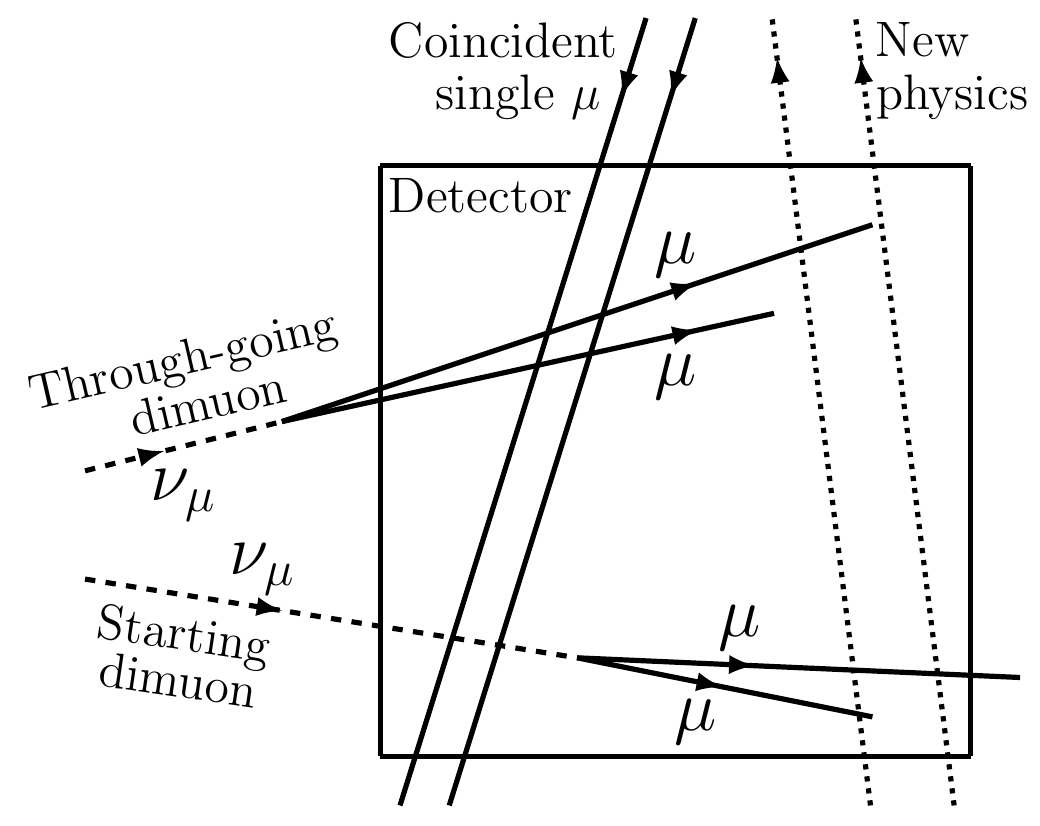}
\caption{
Examples of dimuons and other events with similar signatures.  Dimuon signals may be either starting and throughgoing.  Backgrounds from two coincident single muons are rare except for downgoing events.  New-physics signals, e.g., double staus produced in supersymmetric models~~\cite{Albuquerque:2003mi, Albuquerque:2006am, Ando:2007ds, ICECUBE:2013jjy, Kopper:2015rrp, Kopper:2016hhb, Meighen-Berger:2020eun, IC_stau_2021}, can have a similar topology.
}
\label{fig_dimuons}
\end{figure}

So far, neutrino-telescope analyses are based on studies of just a few event classes, including tracks from muons, showers from other particles, and double bangs from taus.  To get more from the data, new event classes are needed.  Dimuons are an event class that has been detected in accelerator neutrino experiments at energies from tens to hundreds of GeV~\cite{Abramowicz:1982zr, CCFR:1994ikl, CHARMII:1998njb, NOMAD:2000sdx, NuTeV:2001dfo, CHORUS:2008vjb, NOMAD:2013hbk}, but never in neutrino telescopes or in any experiment above 1 TeV.  A dimuon event is two energetic muons produced by the same neutrino.  Dimuon data are important for quantum chromodynamics (QCD), especially for measuring the strange-quark parton distribution function (PDF)~\cite{DeLellis:2004ovi, Hou:2019efy, Faura:2020oom} through deep-inelastic scattering (DIS).   Dimuons can also come from $W$-boson production (WBP)~\cite{Zhou:2019frk}, tridents~\cite{Altmannshofer:2014pba, Magill:2016hgc, Ge:2017poy, Ballett:2018uuc, Gauld:2019pgt, Altmannshofer:2019zhy, Zhou:2019vxt, Zhou:2019frk}, and the Glashow resonance~\cite{Glashow:1960zz, IceCube:2021rpz}.  They are a background for new-physics searches, e.g., where a pair of staus is produced~~\cite{Albuquerque:2003mi, Albuquerque:2006am, Ando:2007ds, ICECUBE:2013jjy, Kopper:2015rrp, Kopper:2016hhb, Meighen-Berger:2020eun, IC_stau_2021}.  Figure~\ref{fig_dimuons} sketches these possibilities.

In this paper, we predict yields for, search for, and discuss the physical significance of neutrino-induced dimuons in high-energy neutrino telescopes.  We focus on the two channels with the largest yields, DIS and WBP.  In Sec.~\ref{sec_pred}, we make our theoretical predictions of dimuon signals, plus backgrounds for them, in IceCube and IceCube-Gen2.  The signals are large enough that IceCube should already have many events, while the backgrounds are mostly either negligible or reducible.  In Sec.~\ref{sec_data}, we search for dimuons in IceCube's 10 years of data for point-source searches~\cite{IceCube:2021xar, data_webpage}, finding 19 dimuon candidates.  Though IceCube has recently shown that these are not real dimuons, the details of our analysis remain valuable, as explained therein.  In Sec.~\ref{sec_dis}, we discuss the exciting physical significance of dimuons, including for measuring the strange-quark PDF at much higher factorization scales than before and possibly for enabling the first detection of WBP.  We conclude in Sec.~\ref{sec_concl}.


\section{Theoretical Predictions}
\label{sec_pred}

In this section, we present our theoretical predictions for dimuons in IceCube and IceCube-Gen2.   In Sec.~\ref{sec_pred_proc}, we discuss the most important neutrino-induced dimuon processes; in Sec.~\ref{sec_pred_calc}, our calculational framework; in Sec.~\ref{sec_pred_rslt0}, we show the cross sections and parent neutrino spectra; in Sec.~\ref{sec_pred_rslt}, our predicted signals; and in Sec.~\ref{sec_calc_Ncoinc}, we discuss our predicted backgrounds.


\subsection{Neutrino-Induced Dimuon Processes}
\label{sec_pred_proc}

The largest neutrino-induced dimuon yield is from charged current neutrino-nucleus DIS~\cite{Gandhi:1995tf, Gandhi:1998ri, CooperSarkar:2011pa, Connolly:2011vc, Chen:2013dza, Bertone:2018dse}, i.e.,
\begin{equation}
\begin{aligned}
\nu_\mu + q_1 (N) \rightarrow \mu^- + q_2 \rightarrow \mu^- + \mu^+ + \nu_\mu + X, 
\label{eq_dimu_DIS}
\end{aligned}
\end{equation}
where the neutrino has energy $E_\nu$, $q_1$ is the incoming quark in the nucleon $N$, $q_2$ the outgoing quark, and $X$ denotes hadrons. The $\mu^-$ comes directly from the leptonic vertex while the $\mu^+$ and $\nu_\mu$ come from the decay of a hadron following the hadronization of $q_2$. (For $\bar{\nu}_\mu$, take a $CP$ transform.)  Theoretically, tau neutrinos could also produce dimuons, with the tau lepton decaying to a muon, but the event rate is negligible due to their low fluxes in atmospheric neutrinos.  Our calculations greatly improve upon the early work in Refs.~\cite{Albuquerque:2003mi, Albuquerque:2006am}, which focused on the new-physics signatures but also made estimates of the DIS backgrounds.

\begin{figure}[t]
\includegraphics[width=0.9\columnwidth]{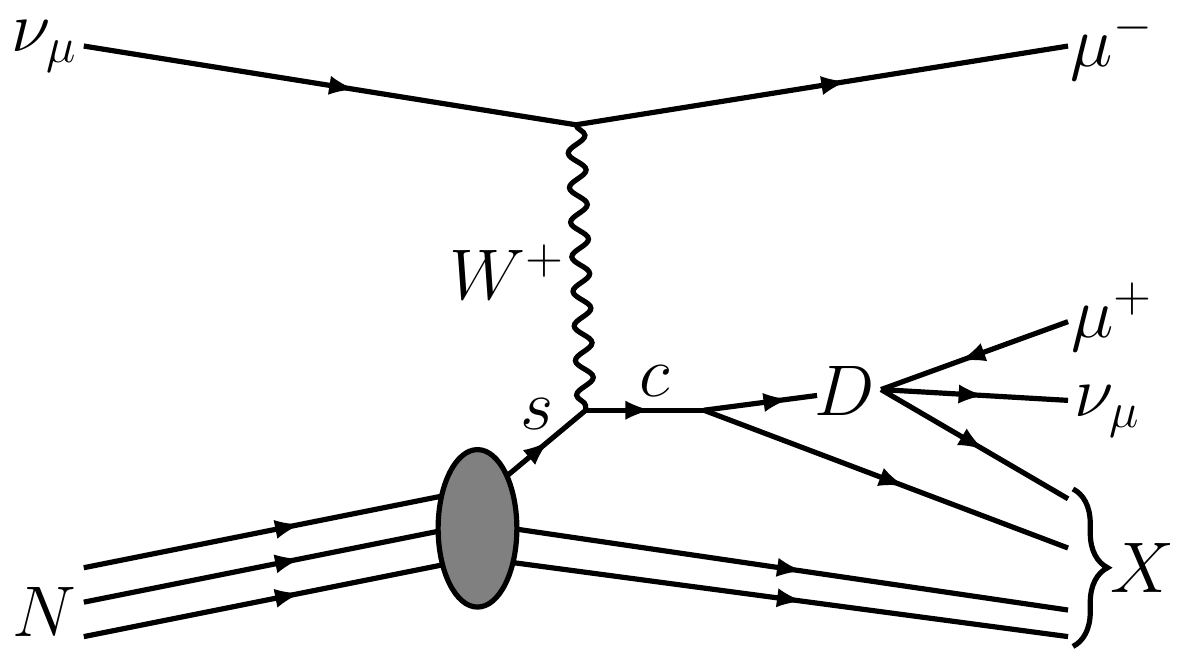}
\caption{
The dominant channel for dimuon production: a strange-quark-induced DIS event between $\nu_\mu$ and a nucleon.  The $D$ meson is from hadronization of the charm quark, and $X$ indicates hadronic particles that produce a shower.  A similar diagram holds for $\bar{\nu}_\mu$.
}
\label{fig_dimu_diagram_DIS}
\end{figure}

Figure~\ref{fig_dimu_diagram_DIS} shows the dominant case of Eq.~(\ref{eq_dimu_DIS}), with an incoming strange quark and an outgoing charm quark.  The charm quark produces a high-energy $D$ meson via hadronization, and the $D$ meson decay products include a high-energy muon.  Our calculation also includes the contributions from incoming down quarks and charm antiquarks, which matter only at high energies.  When the incoming neutrino energy is above a few hundred TeV, there is an important contribution from bottom-quark to top-quark transitions, with $t \rightarrow W \rightarrow \mu$, where $W$ is the W boson~\cite{Barger:2016deu}.  However, the overall event rate is suppressed by the low neutrino flux at high energies. In principle, hadronization produces multiple hadrons that can decay to multiple muons (e.g., trimuon events).  However, the expected event rates are very low and most events would be below detector thresholds.

The next most important dimuon-production process is the neutrino-nucleus WBP~\cite{Seckel:1997kk, Alikhanov:2015kla, Zhou:2019vxt, Zhou:2019frk}.  Though WBP is a significant component of high-energy neutrino interactions (as large as $\simeq 7.5\%$ for ice and $\simeq 14\%$ for iron, which is important for neutrino attenuation in Earth~\cite{Zhou:2019vxt}), it has not yet been identified~\cite{Zhou:2019frk}.  The first calculation of WBP, by Lee and Yang, dates back to the early 1960s~\cite{Lee:1960qv, Lee:1961jj}.  WBP mainly produces dimuons via
\begin{equation}
\begin{aligned}
\nu_\mu + A & \rightarrow \mu^- + W^+ + A'\,,  \text{ with } W^+ \rightarrow \mu^+ \nu_\mu \text{ or } \tau^+ \nu_\tau
\label{eq_dimu_WBP}
\end{aligned}
\end{equation}
where $A$ and $A'$ are the incoming and outgoing nuclei, respectively (the process $W \rightarrow c \rightarrow \mu$ also makes a contribution, but our calculation shows that it is less than 10\%).  The diagrams of this process can be found in Fig.~1 of Ref.~\cite{Zhou:2019vxt}.  The energy transferred to the nucleus is small compared to the energy threshold of high-energy neutrino telescopes~\cite{Zhou:2019vxt, Zhou:2019frk}, so we do not expect muons or hadronic activity from the nuclear side.  (As above, for $\bar{\nu}_\mu$, take a $CP$ transform.) We again ignore tau-neutrino-induced events due to their low fluxes.

Trident events could also produce dimuons~\cite{Altmannshofer:2014pba, Magill:2016hgc, Ge:2017poy, Ballett:2018uuc, Gauld:2019pgt, Altmannshofer:2019zhy, Zhou:2019vxt, Zhou:2019frk}, but they have smaller cross sections and most of them are below the energy threshold of IceCube and especially IceCube-Gen2.  Another possibility is Glashow resonance~\cite{Glashow:1960zz, IceCube:2021rpz}, $\bar{\nu}_e + e^- \rightarrow W^-$, with the $W$ decaying to two quarks that produce two muons through hadronization, but the event rate is highly suppressed due to very low fluxes around $E_\nu \simeq 6.3$ PeV. Note that Glashow-resonance events do not occur for $\nu_e$ in neutrino detectors, due to the lack of positron targets.

Most of our predicted events are opposite-sign dimuons, but same-sign dimuons could occur through higher order processes~\cite{Sandler:1992wj}.  High-energy neutrino telescopes cannot distinguish the charge of a muon.


\subsection{Calculational Framework}
\label{sec_pred_calc}

There are two kinds of dimuon events.  One is starting dimuons, with the parent neutrino interactions occurring in the detector.  The other is throughgoing dimuons, with the neutrino interactions occurring outside the detector, where the two muons travel while losing energy in matter before entering the detector.

We define the more energetic muon as $\mu1$ and the less energetic one as $\mu2$.  We use $E_{\mu1}'$ and $E_{\mu2}'$ to denote their energies at production and $E_{\mu1}$ and $E_{\mu2}$ to denote their energies when entering the detector.  For starting events, $E_{\mu1} \equiv E_{\mu1}'$ and $E_{\mu2} \equiv E_{\mu2}'$.  For DIS, most often $\mu1$ is the muon directly from the leptonic vertex while $\mu2$ is the muon from subsequent hadronization and decay. For WBP, on the contrary, $\mu1$ comes most often from the $W$ decay while $\mu2$ comes directly from the leptonic vertex.  We assume that all $D$ mesons decay before interaction, which would overestimate the DIS dimuon rate above $E_\nu \sim 10$ TeV, where the precise effects are vague due to the cross-section uncertainties~\cite{Barcelo:2010xp, ParticleDataGroup:2020ssz}.  Dimuons from the WBP process, which dominates at such energies, are not affected by this.

The calculation of the spectra of starting dimuons starts with
%
\begin{equation}
\begin{aligned}
& \frac{d N_{\mu\mu}^{\rm st}}{d E_{\mu1/2}} 
\equiv
\frac{d N_{\mu\mu}^{\rm st}}{d E_{\mu1/2}'} 
= N_t T \times  \\
& \ \ \ \ \ 
\int_{E_{\rm th}}^{\infty} d E_\nu
\frac{dF_\nu}{dE_\nu}(E_\nu) 
\frac{d \sigma_{\mu\mu}^\text{cuts}}{dE_{\mu1/2}'} (E_{\mu1/2}', E_\nu | E_{\mu2}'>E_\text{th}) \,,
\end{aligned}
\end{equation}
where $N_t$ is the number of interaction targets (nucleons or nuclei) in the detector (1 $\rm km^3$ of ice in IceCube and 7.9-$\rm km^3$ in IceCube-Gen2), $T$ the observation time, ${d F_\nu}/{dE_\nu}$ the neutrino flux in a certain zenith range after taking into account attenuation in Earth (Sec.~\ref{sec_pred_calc_flux}), $\sigma^\text{cuts}_{\mu\mu}$ the dimuon cross section after an angular-separation cut (Sec.~\ref{sec_pred_rslt0}), and $E_\text{th}$ is the energy threshold of the detector, which is 100 GeV for IceCube~\cite{IceCube_web} and 300 GeV for IceCube-Gen2~\cite{IceCube-Gen2:2020qha}.

For throughgoing dimuons, we develop a framework for such calculations in Appendix~\ref{app_thrgoCalc} that takes into account the energy losses of both muons~\cite{Dutta:2000hh, Groom:2001kq, ParticleDataGroup:2020ssz} and the detector threshold.  The spectrum of throughgoing $E_{\mu2}$ is (Appendix~\ref{app_thrgoCalc}),
\vspace{0.cm}
\begin{equation}
\begin{aligned}
\frac{d N_{\mu\mu}^\text{thr}}{d E_{\mu2}}
& = 
\frac{ A_\text{det} T N_A }{\alpha+\beta E_{\mu2}}  
\int_{E_{\mu2}}^{\infty} dE_\nu \frac{d F_\nu}{dE_\nu}(E_\nu) \\
& \ \ \ \ \ \ \ \ \ \ \ \ \ \ \ \ \ \ \ \ \ \   \int_{E_{\mu2}}^{E_\nu} d E_{\mu2}'
\frac{d \sigma^\text{cuts}_{\mu\mu}}{dE_{\mu2}'}(E_{\mu2}', E_\nu) \,,
\label{eq_main_Nmm_4ints}
\end{aligned}
\vspace{0.0cm}
\end{equation}
where $A_\text{det}$ is the cross-sectional area of the detector (about 1~$\rm km^2$ for IceCube~\cite{IceCube_web} and $(7.9)^{2/3}\ \rm km^2$ for IceCube-Gen2~\cite{IceCube-Gen2:2020qha}), and $N_A = 6.02\times10^{-23} \, \rm g^{-1}$ is the Avogadro number. Here $\alpha = 3.0 \times 10^{-3} \, \rm GeV\, cm^2/g$ and $\beta = 3.0 \times 10^{-6} \, \rm cm^2/g$ characterize the muon energy losses in ice ($dE/dX = -\alpha - \beta E$, where $X$ is the column density), due to ionization and radiation, respectively~\cite{Dutta:2000hh, Groom:2001kq, ParticleDataGroup:2020ssz}.  The variations of $\alpha$ and $\beta$ with energy are slow.

The spectrum of throughgoing $E_{\mu1}$ is (Appendix~\ref{app_thrgoCalc})
\begin{widetext}
\begin{equation}
\begin{aligned}
\frac{d N_{\mu\mu}^\text{thr}}{d E_{\mu1}}
& = 
\frac{A_\text{det} T N_A}{\alpha+\beta E_{\mu1}}
\int_{E_{\mu1}}^{\infty} dE_\nu \frac{d F_\nu}{dE_\nu}(E_\nu)
\int_{E_{\mu1}}^{E_\nu} d E_{\mu1}'   
\int_{E_{\mu2, \rm th}'}^{E_{\mu1}'} d E_{\mu2}' \, 
\frac{d^2\sigma^\text{cuts}_{\mu\mu}}{d E_{\mu1}' d E_{\mu2}'} (E_{\mu1}', E_{\mu2}', E_\nu) \,,
\end{aligned}
\end{equation}
\end{widetext}
where
\begin{equation}
\begin{aligned}
E_{\mu2, \text{th}}' = \left( \frac{E_{\mu1}'+\epsilon}{E_{\mu1}+\epsilon} \right) (E_\text{th}+\epsilon) - \epsilon \,,
\end{aligned}
\end{equation}
and $\epsilon = \alpha/\beta$ is the critical energy (about 1 TeV in ice), above which the radiative energy losses dominate.

For the dimuon cross sections of DIS, we use {\tt MadGraph (v3.1.0)}~\cite{Alwall:2014hca} for the hard processes and {\tt Pythia (v8.305)}~\cite{Sjostrand:2014zea} for the hadronizations and decays.  For WBP, we use the calculational framework we provided in Refs.~\cite{Zhou:2019vxt, Zhou:2019frk}. We use the {\tt CT14qed} PDF set~\cite{Schmidt:2015zda, CT14web}, which provides the inelastic photon, quark, and gluon PDFs self-consistently.

Above the detector thresholds (deposited energies of 100 GeV for IceCube and 300 GeV for IceCube-Gen2), we take the detection efficiency and the muon-track particle-identification efficiency to be unity, which is realistic.  The energy resolution is $\simeq 10\%$, much smaller than the bin sizes we use below, so we neglect smearing.  And the angular resolution is $\simeq 1^\circ$, which is small enough that it only needs to be considered for separating two muons.


\subsubsection{IceCube and IceCube-Gen2}
\label{sec_pred_calc_telescopes}

High-energy neutrino telescopes detect Cherenkov photons emitted by relativistic secondary charged particles produced by neutrino interactions.  A muon forms a long track due to its low energy-loss rate in matter, in contrast to electrons and hadrons, which form a shower that looks like a blob. Compared to showers, tracks have worse energy resolution while better angular resolution, which makes it possible to recognize dimuon events (unlike dielectron events). 

For dimuon detection, the most important aspects of the neutrino telescopes are the angular and energy thresholds, which are determined (see below) by the spacings between the digital optical modules (DOMs) that detect Cherenkov photons.  The DOMs are deployed on the strings in the detector.  We consider two high-energy neutrino telescopes: IceCube~\cite{IceCube_web} and the proposed IceCube-Gen2 (the optical array only)~\cite{IceCube-Gen2:2020qha}.

\begin{figure}[t]
\includegraphics[width=\columnwidth]{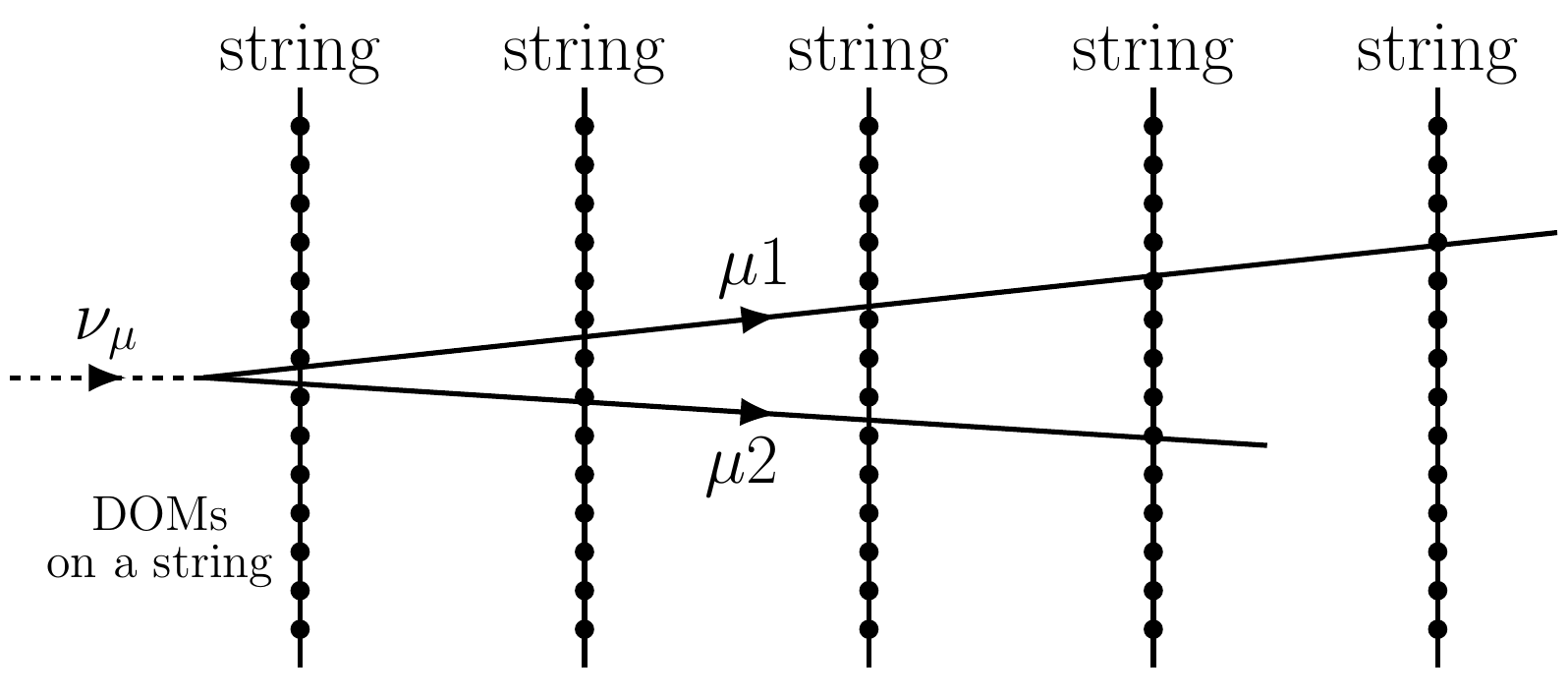}
\caption{
A dimuon event traveling in a high-energy neutrino detector. The vertical lines and the dots on them represent the strings in the detector and the DOMs on them, respectively.  In our calculation, we require $\mu1$ and $\mu2$ to separate by at least $2D_{\rm v}$ to be distinguishable, where $D_{\rm v}$ is the vertical spacing between DOMs (here we show a larger separation).  
}
\label{fig_geometry}
\end{figure}

Figure~\ref{fig_geometry} illustrates a dimuon event in a high-energy neutrino detector.  The horizontal spacing between DOMs is determined by the spacing between the adjacent strings, which is $D_{\rm h} = 125$~m in IceCube (for the 86-string complete configuration, or IC86, which started in 2011)~\cite{IceCube_web}.  For IceCube-Gen2, this is planned to be $D_{\rm h} = 240$~m~\cite{IceCube-Gen2:2020qha}.  The vertical spacing between the DOMs on each string is $D_{\rm v} \simeq 17$~m in IceCube (with 60 DoMs between 1450 and 2450 m below the surface)~\cite{IceCube_web} and planned to be about $D_{\rm v} \simeq 16$~m for IceCube-Gen2 (with 80 DOMs between 1325 m and 2575 m)~\cite{IceCube-Gen2:2020qha}.  

\subsubsection{Angular Separation Cut}
\label{sec_pred_calc_angcut}

To be recognized as being two tracks, dimuons must have a minimum angular separation.  Ultimately, this must be defined by experimental analyses, but we show that a simple criterion captures much of the physics.  We define the angular-separation cut to be (Fig.~\ref{fig_geometry})
\begin{equation}
\begin{aligned}
R_{\mu2} \theta_{\mu\mu} > 2 D_{\rm v} \,,
\label{eq_angcut}
\end{aligned}
\end{equation}
where 
$\theta_{\mu\mu}$ is the angle between $\mu1$ and $\mu2$, and $R_{\mu2}$ is the range of $\mu2$ between its production point and where it exits the detector or stops in it.  For throughgoing muons, we use $R_{\mu2} = \frac{1}{\rho_{\rm ice} \beta} \ln \left[\frac{\alpha+\beta E_{\mu2}'}{\alpha}\right]$, where $\rho_{\rm ice} = 0.92 \, \rm g/cm^3$.  For starting muons, we use the maximum of this range and the detector size (1~km for IceCube or $(7.9)^{1/3}$~km for IceCube-Gen2).

The reason for the angular-separation cut is as follows.   As $\mu1$ and $\mu2$ pass through the detector, their Cherenkov photons trigger nearby DOMs on each string.  Meanwhile, the separation between them increases as they propagate, reaching the maximum of $R_{\mu2} \theta_{\mu\mu}$.  If their separation is large enough to trigger different DOMs on one or multiple strings, they can be identified as dimuons.  As the horizontal DOM spacing is too large to easily distinguish $\mu1$ and $\mu2$, we focus on the vertical spacing, estimating that a threshold of $2 D_{\rm v}$ is reasonable, which seems to be the case.  Strictly speaking, the muons do not hit the DOMs; the light from their Cherenkov cones (angle $\simeq 42^\circ$) does.  We thus refer here to the reconstructed directions.

Our definition is exact for dimuons aligned horizontally in Fig.~\ref{fig_geometry}, such that the two muons trigger different DOMs on the same strings.  For muons aligned perpendicular to the page, detection can still be favorable because there are many rows of strings, and the projected separation between them may still be small.  The most difficult configuration to detect is when the dimuons are aligned vertically, along a string.  However, these events are less important because the atmospheric-neutrino flux peaks near the horizon and because downgoing muons constitute a serious background.  Finally, our angular-separation cut is zenith-angle independent. For a dimuon coming from zenith angles of $\theta_z \neq 90^\circ$, the spacing between the DOMs are effectively smaller, i.e., $D_{\rm v} \sin \theta_z$. However, the widths and separation of two muons being projected vertically is effectively larger, i.e., scaling by $1 / \sin \theta_z$. These two effects cancel.  Therefore, we use Eq.~(\ref{eq_angcut}) for all dimuons.


\subsubsection{Neutrino Fluxes and Earth Absorption}
\label{sec_pred_calc_flux}

The neutrino flux relevant for dimuon production is
\begin{equation}
\begin{aligned}
\frac{d F_\nu}{dE_\nu}(E_\nu)  
=
2\pi \int_{}^{} d \cos\theta_z \frac{d I_\nu}{d E_\nu} (E_\nu, \cos\theta_z) \, e^{-\sigma(E_\nu) C(\theta_z) } \,,
\label{eq_fluxes}
\end{aligned}
\end{equation}
where ${d I_\nu}/{d E_\nu}$ is the neutrino intensity (flux per solid angle).  For atmospheric neutrinos, we use HKKM2015~\cite{Honda:2015fha} (for the South Pole) below 10 TeV and IceCube's measurements~\cite{Borner:2015sed} above 10 TeV.  This flux peaks near the horizon, e.g., being $\simeq 3.5$ times brighter for the horizontal flux compared to the vertical flux at 1 TeV, and even more so at higher energies.  For astrophysical neutrinos, we use IceCube's measurements~\cite{IceCube:2020acn}, though they are not important for dimuon production.  Here $\sigma(E_\nu)$ is the neutrino cross section and $C(\theta_z)$ is the number column density through Earth. We include both DIS and WBP cross sections, as the latter increases the Earth's attenuation by as much as 15\%~\cite{Zhou:2019frk}.  The critical energy for attenuation in Earth is $\simeq 40$ TeV for core-crossing trajectories and is  higher for shorter paths.

\begin{figure*}[t]
\includegraphics[width=0.49\textwidth]{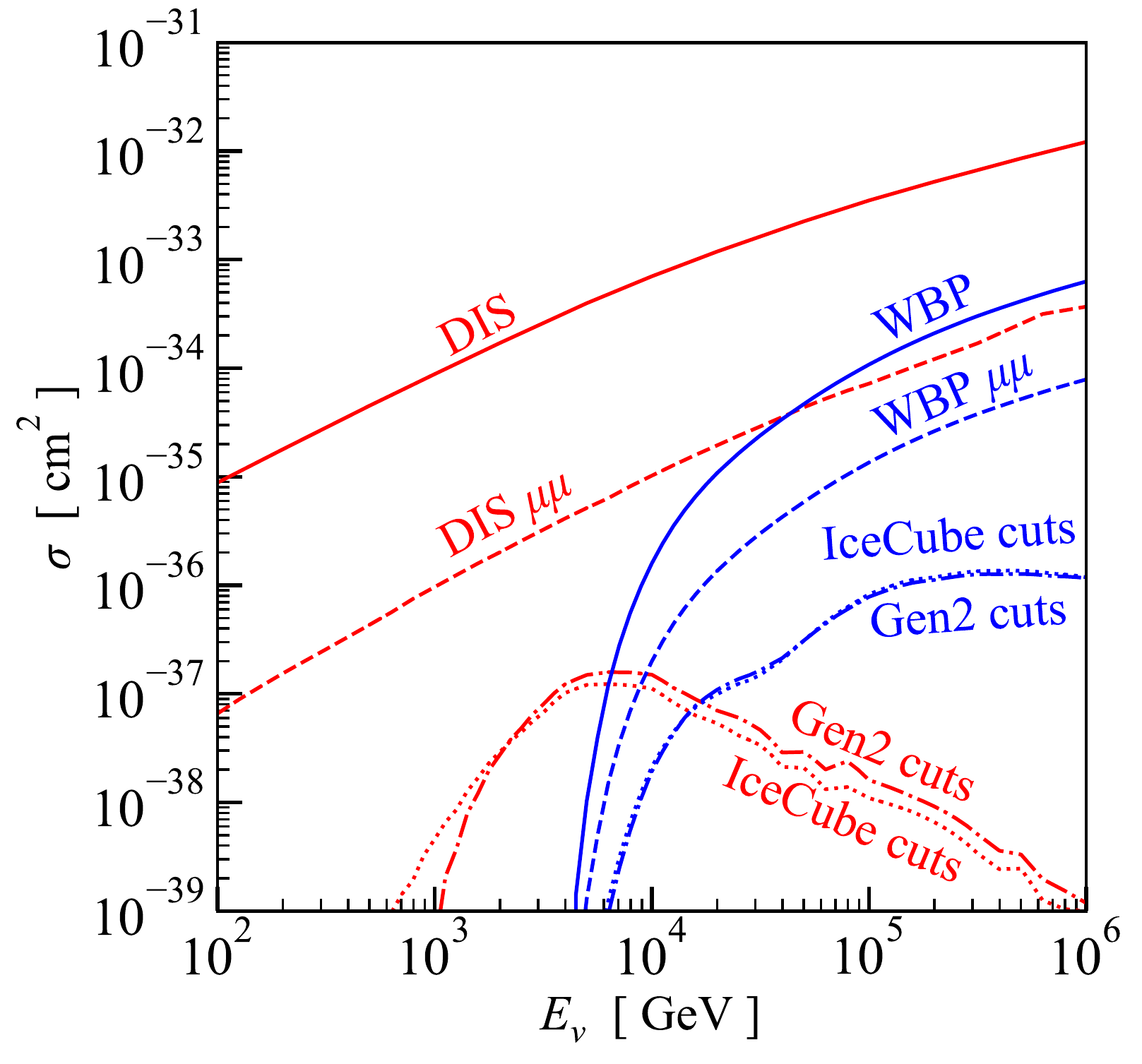}
\includegraphics[width=0.475\textwidth]{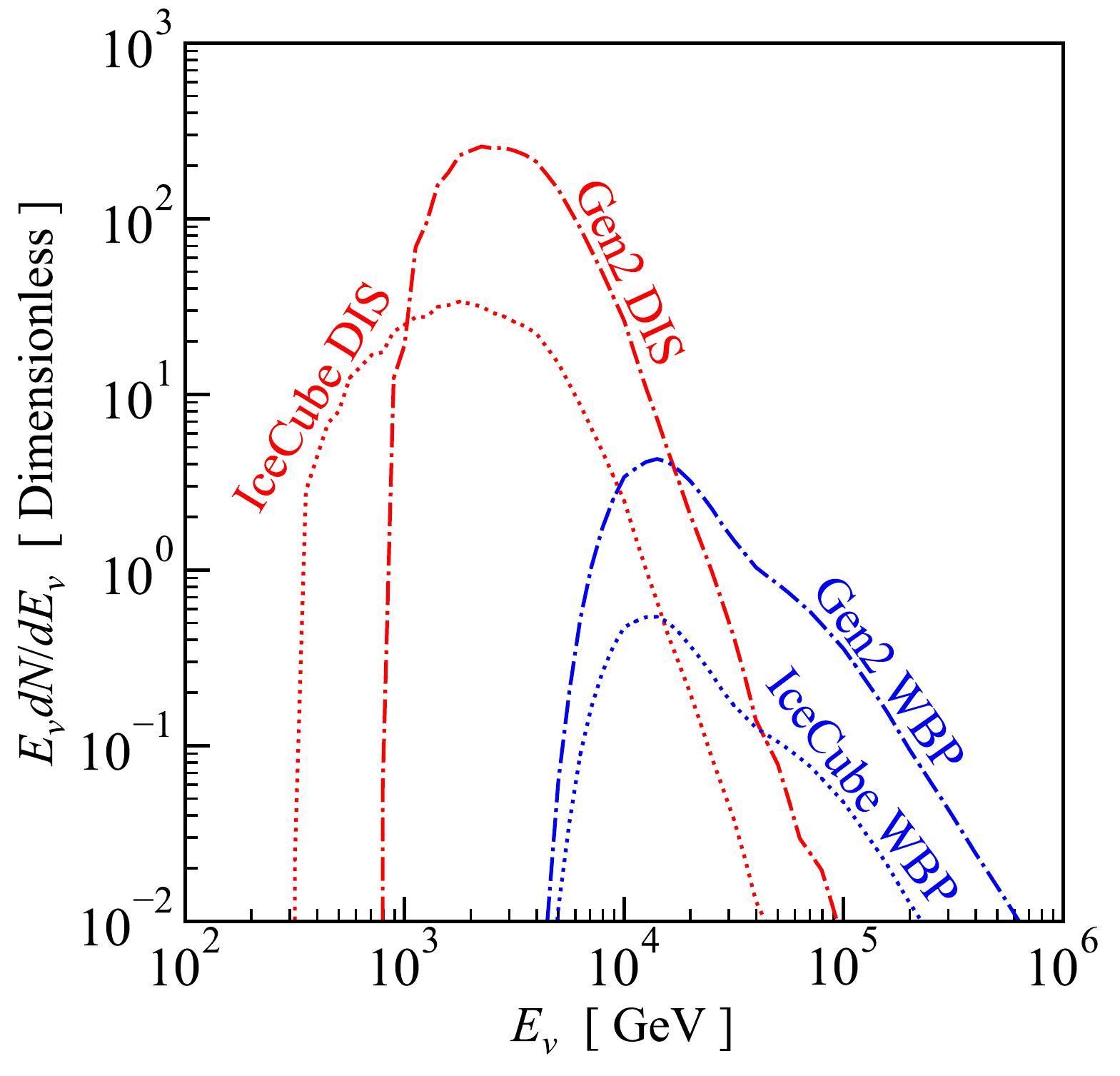}
\caption{
{\bf Left:} Cross sections on $\rm H_2 O$ targets for DIS ({\it red}; $[\nu_\mu+\bar{\nu}_\mu]/2$) from our calculation here and WBP ({\it blue}; $\nu_\mu$ or $\bar{\nu}_\mu$) from our previous work~\cite{Zhou:2019vxt, Zhou:2019frk}.  {\it Solid:} Total cross sections.  {\it Dashed:} Dimuon cross sections.   {\it Dotted:} Dimuon cross sections with cuts of $E_{\mu2} > E_{\rm th}$ and $R_{\mu2} \theta_{\mu\mu} > 2 D_{\rm v}$ for starting muons (Sec.~\ref{sec_pred_calc_angcut}) in IceCube.  {\it Dot-dashed:} same, but for IceCube-Gen2. {\bf Right}: Parent neutrino spectra for starting muons (10 years of exposure; same line styles as the cross sections).  For throughgoing muons, the cut cross sections and parent neutrino spectra (not shown) are more involved but are qualitatively similar.
}
\label{fig_xsec}
\end{figure*}


\subsection{Cross Sections and Parent Neutrino Spectra}
\label{sec_pred_rslt0}

Before calculating the observable dimuon spectra, we first show the dimuon-production cross sections (without and with cuts) and the parent spectra of the neutrinos that produce dimuons.  For simplicity, we focus on starting dimuons.  For throughgoing muons, the results are qualitatively similar but need more involved calculations, which are done in the next subsection.

Figure~\ref{fig_xsec} (left panel) shows the cross sections.  For DIS, the dimuon cross section is $\sim 10^{-2}$ of the total.  (Our calculated DIS cross section is consistent with Refs.~\cite{Gandhi:1995tf, Gandhi:1998ri, CooperSarkar:2011pa, Connolly:2011vc} and our dimuon ratio is consistent with Ref.~\cite{DeLellis:2002bkg, DeLellis:2004ovi} below 1 TeV.)  For WBP, which has a higher threshold due to the heavy $W$ boson, the ratio is given by the corresponding decay branching ratios~\cite{Zhou:2019vxt, Zhou:2019frk}.  

Figure~\ref{fig_xsec} (right panel) shows the parent neutrino spectra, which are the products of the flux (Eq.~(\ref{eq_fluxes})) with the cross sections.  The DIS contribution peaks at $\sim 1$ TeV while the WBP contribution peaks at $\sim 10$~TeV.  These peaks cut off at low energies due either detector (DIS) or production (WBP) thresholds.  They cut off at high energies due to the angular-separation cut, which is less severe for WBP due to the heavy $W$ boson.
%


\subsection{Predicted Dimuon Signals}
\label{sec_pred_rslt}

\begin{table}[b]
\caption{ 
Our predicted numbers of dimuons that could be detected in a full experimental analysis, neglecting backgrounds and other difficulties. The corresponding spectra are shown in Figs.~\ref{fig_dNdE_IceCube} and~\ref{fig_dNdE_Gen2}.   (In Table~\ref{tab_dimu_num_data}, we give the predicted numbers of dimuons for our analysis of public IceCube data.)
}
\label{tab_dimu_num}
\smallskip
\renewcommand{\arraystretch}{1.1} \centering 
\begin{tabular*}{0.4655\textwidth}{c|c|c|c|c}
\hline \hline
             &   \multicolumn{2}{|c|}{Starting} & \multicolumn{2}{|c}{ \hspace{0.0cm} Throughgoing \vspace{0.0cm}}  \\
\hhline{~----}
             & \hspace{0.1299cm} DIS \hspace{0.1299cm}  & \hspace{0.1299cm} WBP \hspace{0.1299cm} & \hspace{0.1299cm} DIS \hspace{0.1299cm} & \hspace{0.1299cm} WBP \hspace{0.1299cm} \\
\hline
IceCube, 10 yrs & 37 & 0.3 & 85 & 6.0 \\
\hline
IceCube-Gen2, 10 yrs & 370 & 5.8 & 231 & 22 \\  
\hline
\end{tabular*}
\end{table}

\begin{figure*}[t]
\includegraphics[width=0.49\textwidth]{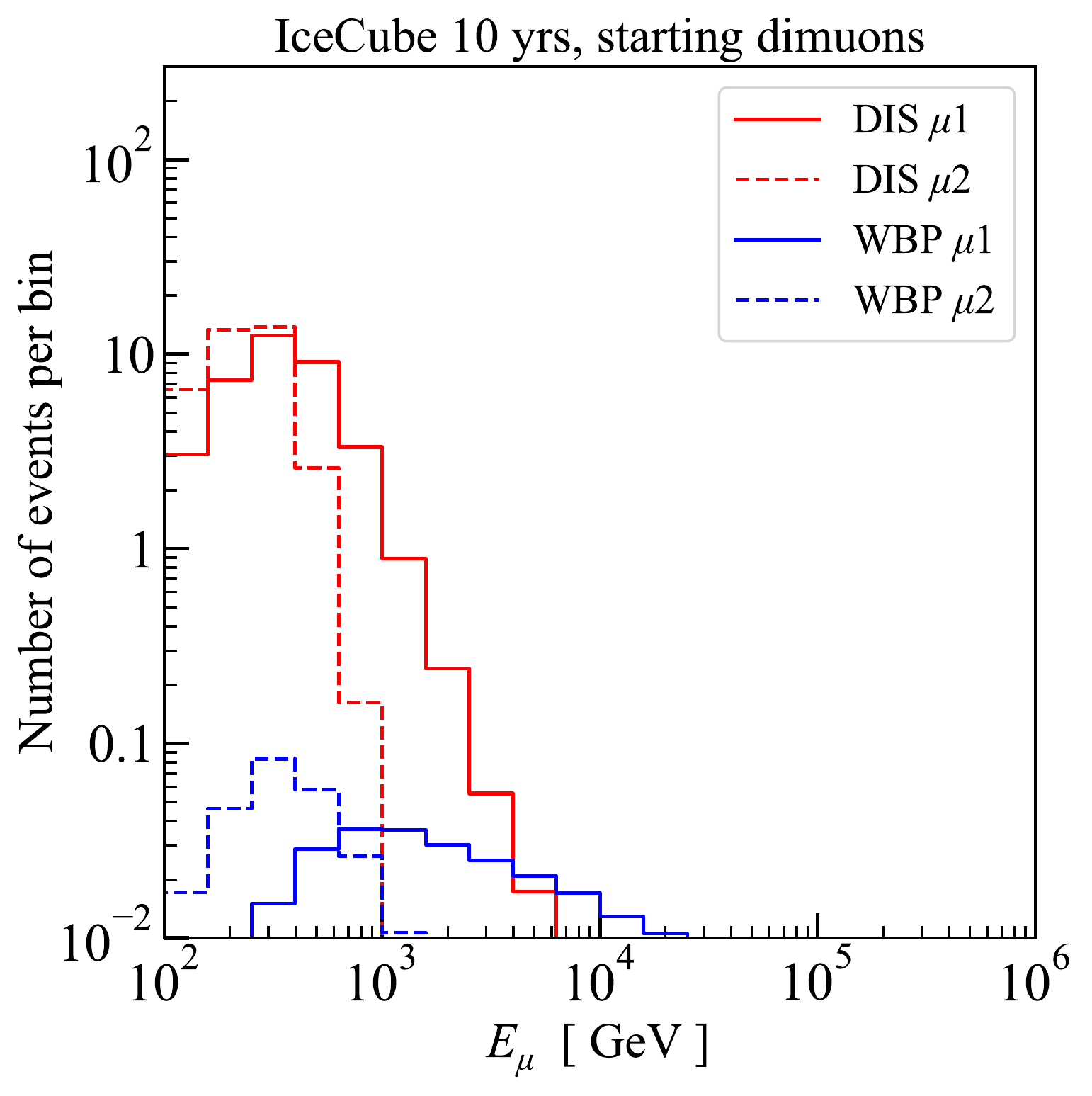}
\includegraphics[width=0.49\textwidth]{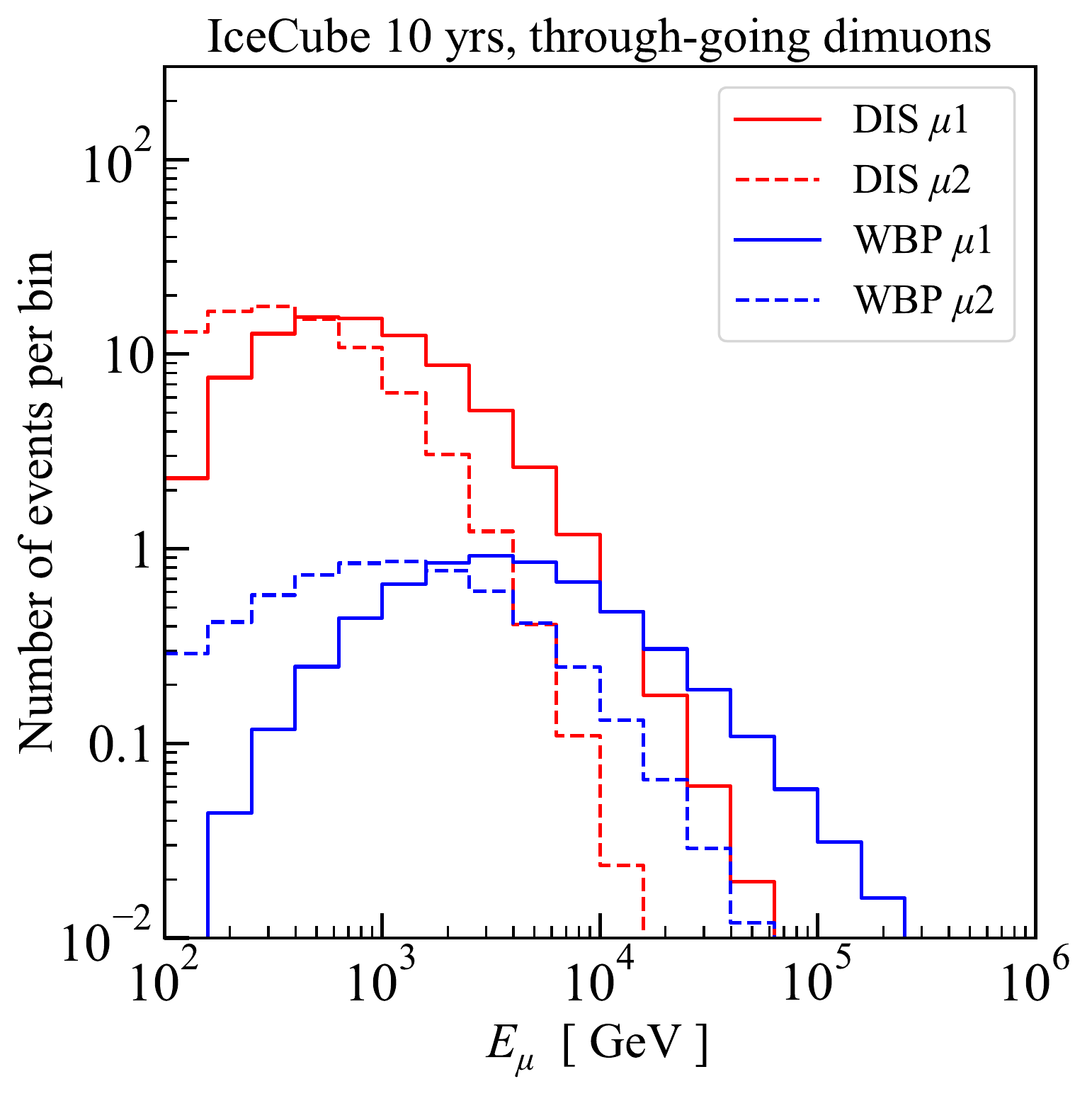}
\caption{
Our predicted dimuon spectra in IceCube.  {\bf Left:} Starting dimuons ($\simeq 37$ from DIS and $\simeq 0.3$ from WBP).  {\bf Right:} Throughgoing dimuons ($\simeq 85$ from DIS and $\simeq 6.0$ from WBP).  We define $\mu1$ to be the more energetic of the two muons.
}
\label{fig_dNdE_IceCube}
\end{figure*}

\begin{figure*}[t]
\includegraphics[width=0.49\textwidth]{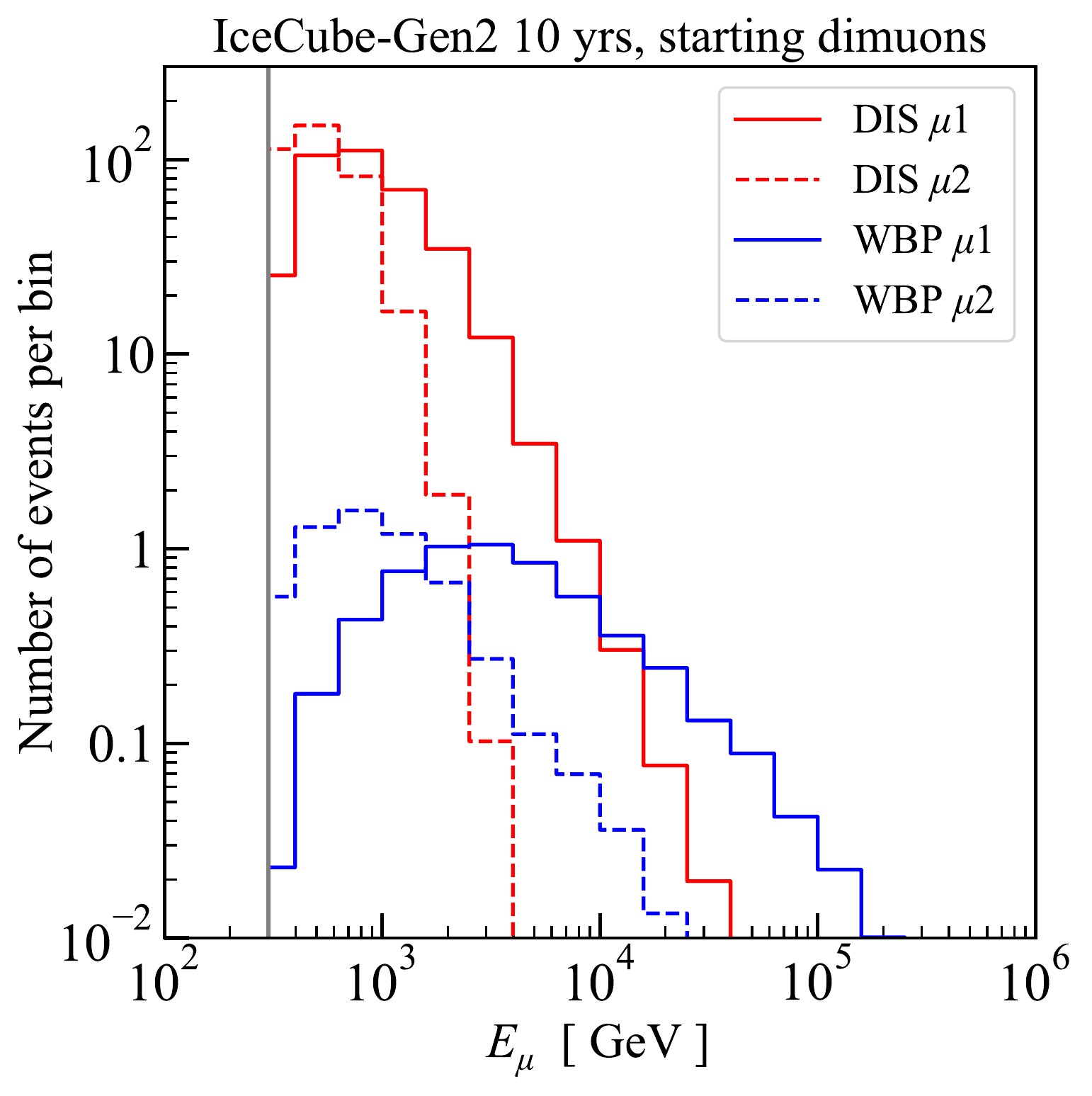}
\includegraphics[width=0.49\textwidth]{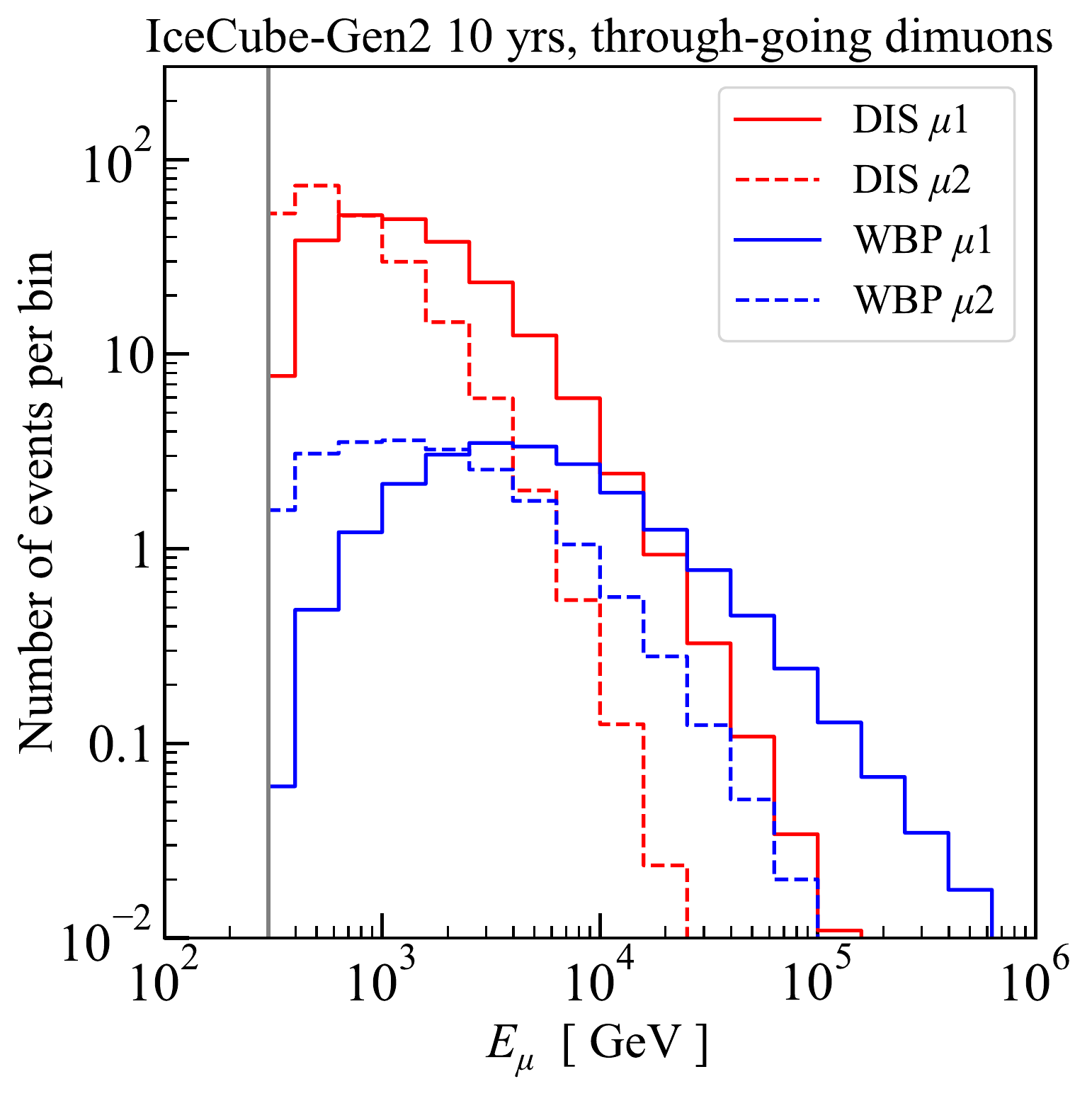}
\caption{
Same as Fig.~\ref{fig_dNdE_IceCube}, but for 10 years of the IceCube-Gen2 optical array.  {\bf Left:} Starting events ($\simeq 370$ from DIS and $\simeq 5.8$ from WBP).   {\bf Right:} Throughgoing events ($\simeq 230$ from DIS and $\simeq 22$ from WBP). 
}
\label{fig_dNdE_Gen2}
\end{figure*}

Figures~\ref{fig_dNdE_IceCube} and \ref{fig_dNdE_Gen2} show our predicted dimuon spectra for IceCube and IceCube-Gen2, respectively, taking into account all effects.  Importantly, we display our results in terms of detectable muon energy (for starting muons, the initial energy; for throughgoing muons, the detector-entering energy), making them directly comparable to experimental data.  The energy of a muon can be measured by its energy-loss fluctuations (for higher energies) or its range (for lower energies)~\cite{IceCube:2021oqo}.  In ice, the ranges of muons of energy 0.1, 1, and 10 TeV are about 0.4, 2.5, and 8.5 km, respectively.  For energies at a few to several hundred GeV, measuring the muon energy can be less precise, which will smear the peaks.  Table~\ref{tab_dimu_num} summarizes the total numbers of events for different channels.  Both throughgoing and starting events are important; in IceCube-Gen2, which has a higher threshold, starting events become more important.  Almost all of the events are from atmospheric neutrino interactions, with astrophysical neutrinos contributing less than $\simeq 3\%$.

These figures show some key features.  The DIS spectra are large, dominating the WBP spectra.  For both, the spectra of $\mu2$ are softer than those of $\mu1$, but they are close enough that both muons can be measured.  Comparing to Fig.~\ref{fig_xsec} (right panel), which is for starting muons only, WBP events carry smaller fractions of the neutrino energy.  Comparing the left and right panels of both Figs.~\ref{fig_dNdE_IceCube} and \ref{fig_dNdE_Gen2}, throughgoing events dominate at the highest muon energies, which is expected due to the long muon range at high energies.  Finally, the separation of DIS and WBP events is more challenging than it appears in Fig.~\ref{fig_xsec} (right panel).  This is primarily due to the different kinematics of DIS and WBP interactions.  In Sec.~\ref{sec_dis_WBP}, we discuss how to separate them, exploiting the lack of hadronic showers in WBP events~\cite{Zhou:2019vxt, Zhou:2019frk}.  

We estimate the overall theoretical uncertainties in our predictions. For the atmospheric neutrino flux, the uncertainty at the relevant energies (Fig.~\ref{fig_xsec}) is about $25\%$~\cite{Honda:2015fha, Borner:2015sed, Schoneberg:2016rkg}. 
For DIS cross sections and hadronization, the overall uncertainty is about $10\%$ at somewhat lower energies~\cite{Faura:2020oom, Dulat:2015mca}, though it would mildly increase with energy.
For WBP, the uncertainty is about $15\%$~\cite{Zhou:2019vxt}.
Combining the flux and interaction uncertainties in quadrature, the overall theoretical uncertainty of our predictions is estimated to be about 30\% for both DIS and WBP dimuons.  This is large enough to cover the uncertainties due to the variation of $\alpha$ and $\beta$ with energy, etc.


\subsection{Dimuon Backgrounds}
\label{sec_calc_Ncoinc}

A background for dimuon signals is two coincident single-muon events.  Our conservative calculations below show that this background is negligible for all-sky starting dimuons and for northern-sky throughgoing dimuons, while perhaps reducible for southern-sky throughgoing dimuons.  Atmospheric muon bundles~\cite{Gaisser:1985yw, FREJUS:1989lko}, i.e., causal associations of muons, are relevant only for southern-sky throughgoing dimuons.  Other possible backgrounds are discussed below.

For two independent muons to mimic a dimuon, they must appear within the same small time ($\delta_T$), angular ($\delta_\theta$), and positional ($\delta_D$) intervals.  The probability of such a coincidence is $\left[ R_\mu \delta_T \delta_\theta^2 \delta_D^2/(A_{\rm tot} A_\text{det}) \right]^2$, where $R_\mu$ is the all-sky single-muon rate, $A_{\rm tot} \simeq 41,253 \ \rm deg^2$ is all-sky angular area, and $A_\text{det}$ is the area of the detector.  Then, the number of all-sky coincident muons in an observation time $T$ is 
\begin{equation}
N_{\mu\mu}^{\rm coinc} = 
\left[ \frac{R_\mu \delta_T \delta_\theta^2 \delta_D^2}{A_{\rm tot} A_\text{det}} \right]^2
\frac{T A_{\rm tot} A_\text{det}}{ \delta_T \delta_\theta^2 \delta_D^2} 
=
R_\mu^2 T^2 \frac{\delta_T}{T} \frac{\delta_\theta^2}{A_{\rm tot}} \frac{\delta_D^2}{A_\text{det}} \, ,
\label{eq_Ncoinc}
\end{equation}
with the last three terms being temporal, directional, and positional, respectively.  For starting dimuons, there should also be a term for coincidence on the depth position; we ignore this as the backgrounds for starting dimuons are already small.

The $\delta_T$ is set by the timing resolution of the detector for track events.  For both IceCube and IceCube-Gen2, an optimistic choice is the timing resolution of the DOMs, which is a few ns.  (For real dimuon events, the muons are simultaneous.)  A conservative choice would be the time it takes a high-energy muon to cross the detector, $\sim 3 \times 10^{-6}$~s for IceCube.  However, the IceCube data we use~\cite{IceCube:2021xar, data_webpage} in Sec.~\ref{sec_data} has time bins of $\simeq 8.6 \times 10^{-4}$ s, though this is a choice, not a detector limitation.  To be conservative, we adopt this value, though IceCube can significantly improve upon it.  The $\delta_\theta$ is set by the maximal $\theta_{\mu\mu}$, which we find to be $\theta_{\mu\mu}^{\rm max} \simeq 5^\circ$ for our predicted events, so we use $\delta_\theta = 10^\circ$ to be conservative, much larger than detector angular resolution for tracks ($\simeq 1^\circ$).  The $\delta_D$ is set by $R_{\mu2}^{\rm max} \theta_{\mu\mu}^{\rm max}$, which we conservatively take to be a few kilometers times $5^\circ$, obtaining a few hundred meters, which is larger than the position resolution of the detector.

We first consider backgrounds for all-sky starting dimuons.  The relevant single muons are almost all induced by atmospheric neutrinos.  For starting single muons, the all-sky event rate in IceCube is $R_\mu \simeq 0.02\, \rm s^{-1}$ for $E_\nu > 100$~GeV.  So conservatively, Eq.~(\ref{eq_Ncoinc}) gives $ N_{\mu\mu}^{\rm coinc} \simeq 3\times10^{-2}$ for 10 years of IceCube.  For 10 years of IceCube-Gen2 ($E_\nu > 300$~GeV), $N_{\mu\mu}^{\rm coinc} \simeq 4\times10^{-2}$.  For northern-sky throughgoing single muons, the rate $R_\mu$ is $\simeq 10$ times larger in IceCube and a smaller increase for IceCube-Gen2, meaning $N_{\mu\mu}^{\rm coinc}$ is at most $\simeq 100$ times larger than those from starting single muons.  All of these background rates, even when estimated conservatively, are negligible.

For southern-sky throughgoing dimuons, the relevant single tracks are atmospheric muons.  The total rate is $R_\mu \simeq 1500\ \rm s^{-1}$ for IceCube and $\simeq 7.9$ times larger for IceCube-Gen2~\cite{IceCube:2015wro}.  For $T=10$~years, Eq.~(\ref{eq_Ncoinc}) gives $N_{\mu\mu}^{\rm coinc} \simeq 2\times10^8$ (IceCube) and $4\times10^9$ (IceCube-Gen2), which are both huge.  However, dedicated experimental analyses may be able to reduce these backgrounds by orders of magnitude and also lower the energy threshold due to the temporal, directional, and positional correlations of $\mu1$ and $\mu2$.  There are several reasons.  True neutrino-induced events would be produced in the ice, $\lesssim 1.5$ km above the detector, compared to coincidences of atmospheric muons or muon bundles, for which the muons are produced $\sim$ 10--20 km above the detector.  The IceTop detector~\cite{Tosi:2019nau, Tosi:2017zho, Tosi:2015bhm} can be used as a powerful veto.  It would be realistic to use a much smaller time interval~\cite{IceCube_tres1, IceCube_tres2} while keeping all of the signal.  Using a smaller angular interval would greatly reduce backgrounds while only somewhat reducing the signal.  Raising the energy threshold would help, because the spectra are falling and the background coincidence rate depends on the spectrum squared.  Additionally, dimuons have have less stochasticity in their energy losses compared to single muons (this technique has been used to reject downgoing muon bundles~\cite{IceCube:2012iea, IceCube:2021aen}).


\section{Observational Results}
\label{sec_data}

In this section, we present our observational results on dimuon candidates in IceCube public data. 
In Sec.~\ref{sec_data_search}, we explain the dataset and how we search.  In Sec.~\ref{sec_data_positive}, we discuss why these events appear to be real dimuons, followed by Sec.~\ref{sec_data_negative}, where we discuss the arguments against this.

Subsequent to our paper appearing, visual inspection of these events by the IceCube Collaboration reveals that they are not real dimuons, but instead arise from an internal reconstruction error that identifies some single muons crossing the dust layer as two separate muons.  

To help IceCube and the broader community, we include the updated full details of our analysis.  The first reason is to help future dimuon searches, showing what works and what needs further attention.  The second reason is to help identify a subtle background that could affect other searches, e.g., a point-source search (especially transients) that finds two associated events versus one would be quite different.  The third reason is to model a productive exchange between theorists and an experimental collaboration, where IceCube provided helpful feedback before and after our paper appeared, and where we shared this feedback publicly.


\subsection{Dataset and Search for Dimuons}
\label{sec_data_search}

Almost all IceCube data is private, but the collaboration has released a limited dataset covering 10 years (April 2008 to July 2018) of all-sky events~\cite{IceCube:2021xar, data_webpage}.  This dataset is optimized for point-source searches and is intended to facilitate multimessenger analyses.  There are 1,134,450 muon-track events, which are obtained after multiple strong cuts to reduce backgrounds.  The energy distribution of this dataset is strongly suppressed by cuts below a muon energy of about 400 GeV, so we take this as an approximate threshold.  The transition in the spectrum is broad, so the precise threshold is uncertain, which leads to a factor-of-two uncertainty in the DIS yields and much less sensitivity in the WBP yields. We thus expect to find fewer dimuons than our prediction in Sec.~\ref{sec_pred_rslt}.  

For these events, the arrival times are given in the unit of modified Julian day (MJD), downsampled to a precision of $1 \times 10^{-8}$ day ($8.6 \times 10^{-4}$ s), much longer than any physically relevant timescale.  The energy uncertainties are not given.  The angular distributions  are consistent with a uniform distribution in right ascension and a reasonable distribution in declination (see Figs.~1 and 2 of Ref.~\cite{Zhou:2021rhl}).  The 1$\sigma$ angular errors of the events are typically $\lesssim 1^\circ$.  The event positions are not given.  Despite the limitations of this dataset, it is adequate for our dimuon search.  With the greater details available to a collaboration analysis, especially the ability to create a dataset optimized for dimuon searches instead of point sources, much better results are expected.

We search for dimuons in the dataset through the following procedure.  First, we sort the events in terms of their arrival times.  Second, we group time-adjacent events into all possible pairs and sort the pairs by the differences in their arrival times.  Third, we search for pairs with the same arrival times (within $8.6 \times 10^{-4}$ s).  In total, we find 21 such pairs, which already gives a huge reduction in the number of events.  Finally, we check the angular distance between the muons in each pair.  Only two pairs have large angular distances ($\simeq 50^\circ$ and $\simeq 60^\circ$, much larger than the $\theta_{\mu\mu}^{\rm max} \simeq 5^\circ$ expected for dimuons).  Thus, these two pairs are only temporally coincident and are backgrounds (as discussed below, these are consistent with expectations).  For the other 19 pairs, their angular distances are mostly smaller than $\theta_{\mu\mu}^{\rm max}$, except two pairs with $\simeq 9^\circ$ but also with large angular uncertainties.  These 19 pairs thus appear to be dimuons.  Following a similar procedure, we do not find trimuons or other multimuon candidates.  Appendix~\ref{app_dimu} gives further details on these 19 dimuon candidates.

\begin{figure*}[t!]
\includegraphics[width=0.99\columnwidth]{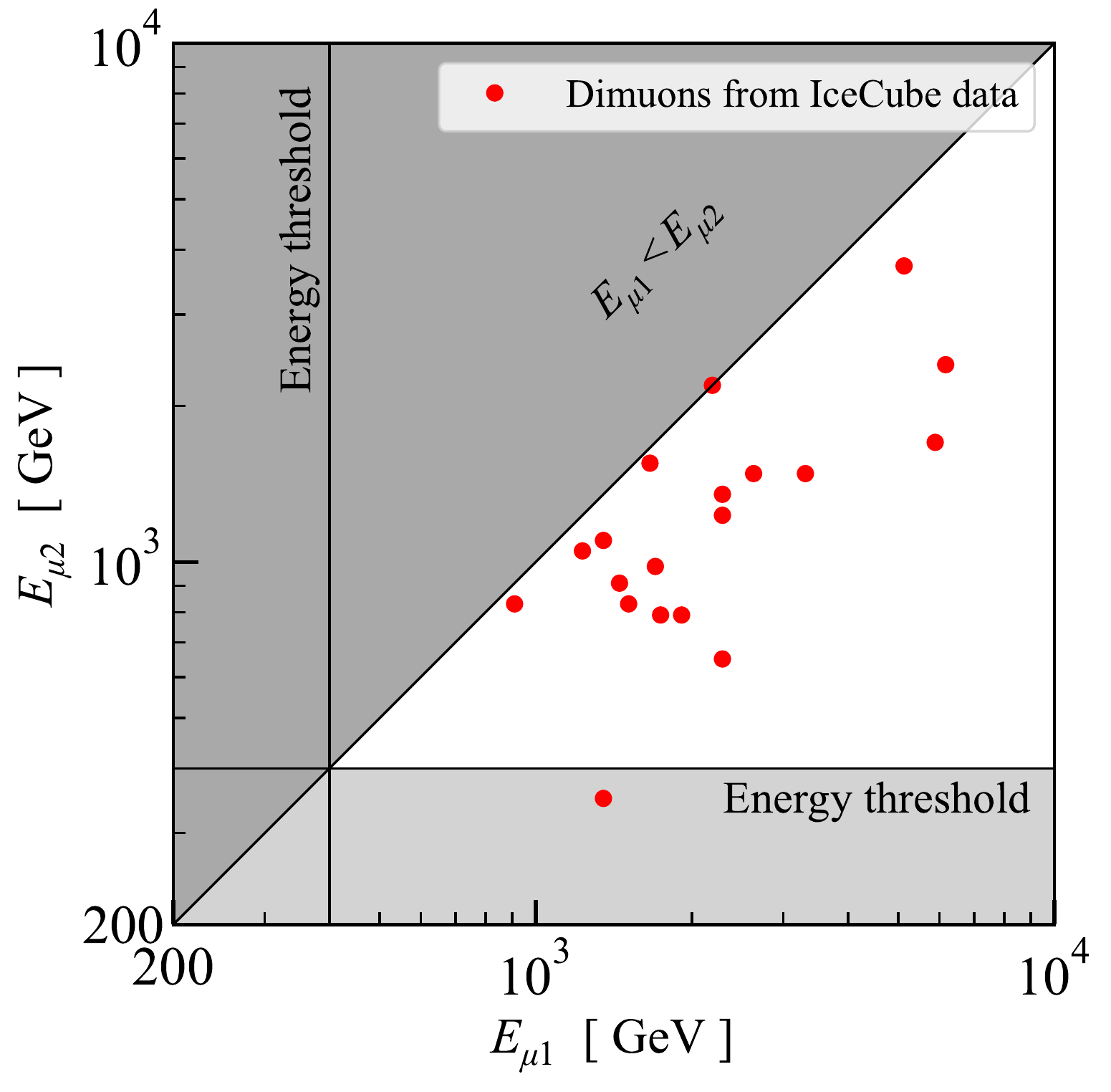}
\includegraphics[width=0.985\columnwidth]{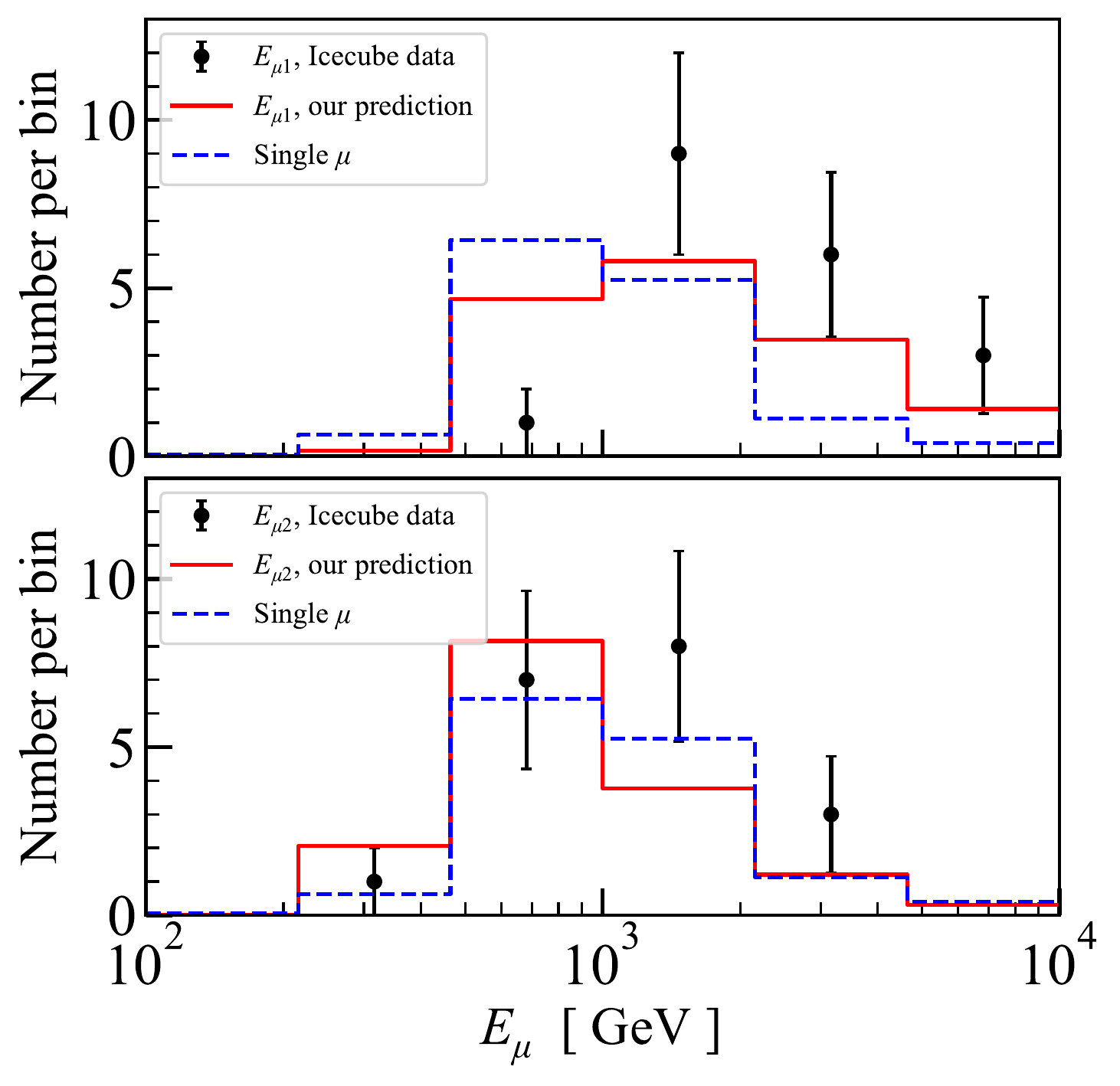}
\caption{
Energy distributions of the 19 dimuon candidates from the IceCube data.  {\bf Left:} Scatter plot of the energies for each muon in the dimuon event. The gray regions are disfavored by either energy threshold (we estimate 400 GeV from the energy distribution of muon events in the IceCube dataset) or the definition $E_{\mu1} \geq E_{\mu2}$.  {\bf Right:} Binned distributions of $E_{\mu1}$ ({\bf upper}) and $E_{\mu2}$ ({\bf lower}), compared to our prediction (solid line) as well as the distribution of single-muon events, as if the dimuon candidates were somehow misreconstructed single muons. See text for details.
}
\label{fig_dimu_data}
\end{figure*}


\subsection{Why These Events Appear to Be Dimuons}
\label{sec_data_positive}

We first examine the basic characteristics of the 19 candidate dimuons.  The arrival times of these events are all after about May 2011, when IceCube's construction was completed (86 strings).  Between about April 2008 and May 2011, the incomplete IceCube detector had a smaller acceptance, especially at lower energies.  In the following, we focus on the period with the full detector (May 2011 to July 2018~\cite{IceCube:2021xar, data_webpage}, 6.93 years live time), which will reduce the yield relative to our predictions.  Within this, the arrival-time distribution is apparently uniform.  The energy distribution is comparable to that of single-muon events, with some important differences, discussed below.  The right-ascension distribution is apparently uniform.  (Appendix~\ref{app_dimu} has figures that show the right-ascension and zenith angular distributions.)  For the zenith angles ($\rm = Dec + 90^\circ$ for IceCube's location) of the dimuon candidates, they are all northern-sky events (upgoing), indicating that they are not atmospheric muons.  In fact, all candidates are at $\rm Dec \gtrsim 10^\circ$, below the horizon. The dataset has strong cuts to reduce backgrounds, which suppresses southern-sky events.  The angular distribution favors horizontal over vertically upward events.  All of these characteristics are consistent with atmospheric-neutrino events.

\begin{table}[b!]
\caption{ 
Our predicted numbers of dimuons expected in our analysis of public IceCube data, taking into account realistic factors.  The corresponding spectra are shown in Fig.~\ref{fig_dimu_data}, and the angular separations are summarized in Appendix~\ref{app_dimu} and shown in Fig.~\ref{fig_dimu_dNdtheta}.
}
\label{tab_dimu_num_data}
\smallskip
\renewcommand{\arraystretch}{1.1} \centering 
\begin{tabular*}{0.4673\textwidth}{c|c|c|c|c}
\hline \hline
             &   \multicolumn{2}{|c|}{Starting} & \multicolumn{2}{|c}{ \hspace{0.0cm} Throughgoing \vspace{0.0cm}}  \\
\hhline{~----}
             & \hspace{0.18cm} DIS \hspace{0.18cm}  & \hspace{0.18cm} WBP \hspace{0.18cm} & \hspace{0.18cm} DIS \hspace{0.18cm} & \hspace{0.18cm} WBP \hspace{0.18cm} \\
\hline
\hspace{0.009cm} IceCube, 6.93 yrs \hspace{0.009cm} & 0.95 & 0.04 & 13.0 & 1.6 \\
\hline
\end{tabular*}
\end{table}

The expected number of background events from coincident single muons is negligible, following the approach of Sec.~\ref{sec_calc_Ncoinc}.  With 897,406 muon-track events in $T \simeq 6.93$ years of exposure~\cite{IceCube:2021xar, data_webpage}, we obtain an event rate (for the cuts of this dataset) of $R_\mu \simeq 4\times10^{-3} \, \rm s^{-1}$.  For the temporal interval, we use $\delta_T \simeq 8.6\times10^{-4} \, \rm s$.  For the directional interval, we estimate $\delta_\theta \simeq 3^\circ$ from the data in Table~\ref{tab_data_dimu_list}.  The positional data are not public, so we take $\delta_D^2 / A_{\rm det} = 1$.  Then, from Eq.~(\ref{eq_Ncoinc}), we expect only $\simeq 6.9 \times 10^{-4}$ coincident single muons in this dataset.

To better understand the backgrounds, we also consider temporal-only coincidences, neglecting the angular information.  The expected number of such events between May 2011 and July 2018 is $\simeq 3.2$ from Eq.~(\ref{eq_Ncoinc}) with the directional and positional terms ignored and input numbers taken directly from the dataset~\cite{IceCube:2021xar, data_webpage}.  As noted in the previous subsection, we find two such events, which we discard due to their huge separation angles.  We also examine muon pairs with fixed arrival-time differences larger than zero.  As expected, we find that the muons in each pair come from random directions of the whole sky.  In sum, our estimates of the backgrounds are in good agreement with expectations.

Last but not least, we compare the 19 dimuon events with our theory, beginning with the yield.  With a threshold of 400 GeV, we predict $\simeq 14$ DIS and $\simeq 1.6$ WBP dimuons (mostly throughgoing), which is in excellent agreement with our observed dimuon candidates, taking into account the $\simeq 30\%$ theoretical uncertainties (Sec.~\ref{sec_pred_rslt}) and the factor-of-two experimental uncertainties for the point-source dataset due to this threshold effect (Sec.~\ref{sec_data_search}).
The details are summarized in Table~\ref{tab_dimu_num_data}.  The most important factor for reducing the prediction from $\simeq 130$ events is the increased energy threshold.  This effect can be estimated from Fig.~\ref{fig_dNdE_IceCube}.  We also take into account the reduction in exposure from 10 years and the reduction in solid angle from $4 \pi$.

Figure~\ref{fig_dimu_data} (left panel) shows the 19 candidate dimuons in the plane of $E_{\mu1}$ and $E_{\mu2}$.  Almost all the events are within the white region, as expected.  One event is detected with a lower value of $E_{\mu2}$, but this is plausible, as the detector would have already been triggered by the deposition of $E_{\mu1}$ (or it might just be energy resolution smearing).

\begin{figure}[h!]
\includegraphics[width=0.92\columnwidth]{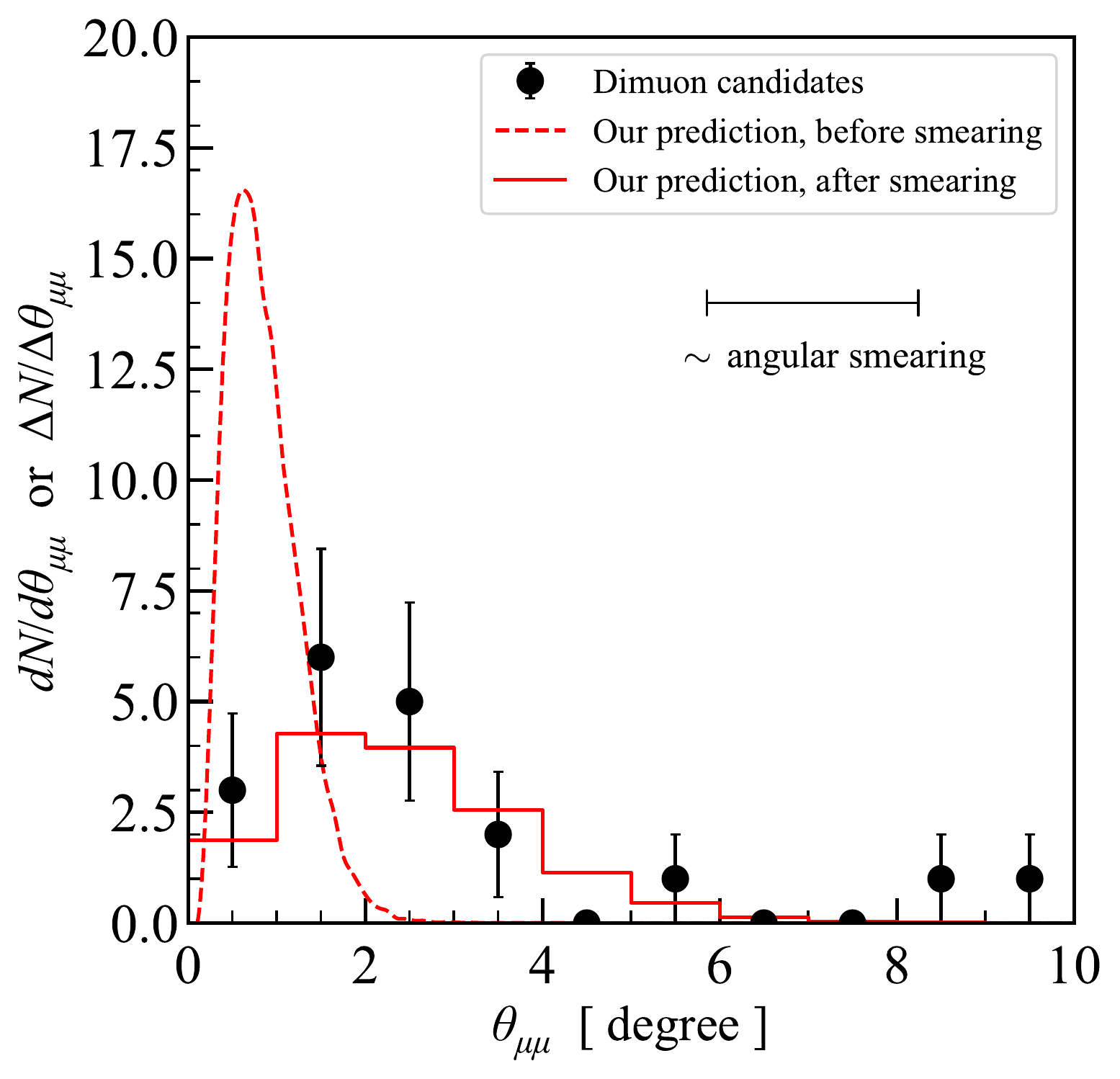}
\caption{Angular-separation distribution of the 19 dimuon candidates and our prediction, which is dominated by throughgoing DIS events (see Table~\ref{tab_dimu_num}).  The dropoff at low $\theta_{\mu\mu}$ is primarily due to the angular cut (see Sec.~\ref{sec_pred_calc_angcut}), while the dropoff at large $\theta_{\mu\mu}$ is caused by the kinematics, which are affected by the energy threshold.  We show results without and with angular-resolution smearing (see text); we show a representative scale for the effective uncertainty.}
\label{fig_dimu_dNdtheta}
\end{figure}

Figure~\ref{fig_dimu_data} (right panel) shows the energy distributions for both the data and our theoretical prediction. We also show the energy distribution of single-muon events, which is directly obtained from the dataset and rescaled to include 19 events, as if the dimuon candidates were somehow misreconstructed single muons.  The data is clearly inconsistent with the single-muon prediction (for example, in the bin of $E_{\mu1} = 3$ TeV, there are 6 dimuons observed but only $\simeq 1 \pm 1$ expected), so we do not show the uncertainties on the single-muon prediction.  Instead, to show how well the spectrum of dimuon events has been measured, we show the Poisson uncertainties on the data.  The shape of the dimuon spectrum is in excellent agreement with the data.

Figure~\ref{fig_dimu_dNdtheta} shows the angular-separation distribution of the data compared to our prediction, finding good agreement.  The normalization of our prediction is absolute (taking into account exposure time, threshold, etc.), and is not scaled to 19 events.  The prediction before angular smearing (red dashed) is calculated directly using Eq.~(\ref{eq_app_dNdtheta}).  To take into account angular smearing, we simulate a large number of $\mu1$ and $\mu2$ events with angular separations randomly drawn from the red dashed curve. Then for each of $\mu1$ and $\mu2$, we smear their directions separately with 2-d symmetric normal distributions with the standard deviations randomly picked from Table~\ref{tab_data_dimu_list} (AngErr1 and AngErr2 columns).  Finally, we calculate the new angular separation and histogram the results.  (We obtain similar results if we treat all the angular uncertainties as being $2.5^\circ$.)  Two events are observed at large separation angles, where we predict none.  However, the discrepancy is much less than it might seem.  First, these two points have especially large angular uncertainties, and are each only $\sim 2 \sigma$ sideways from the bulk of the distribution.  Second, the IceCube angular-smearing distributions are known to have long tails to high separations~\cite{IceCube:2014gax}, which we do not take into account, so the true deviations are less significant. 

In conclusion, the 19 candidate dimuon events appear consistent with being signal events.


\subsection{Why These Events Are Not Dimuons}
\label{sec_data_negative}

As noted, the IceCube Collaboration finds after visual inspection that these candidates are not real dimuons, but instead arise from single muons misreconstructed as two muons.  This arises because of the difficulty of reconstructing muon tracks that cross the dust layer at a depth of 2000--2100 m, which causes significant light scattering and absorption.  IceCube has analysis tools to cope with this, but these failed in a small fraction of cases.

Before this definitive result from the IceCube Collaboration, we received skeptical remarks about our candidate events from IceCube experts~\cite{pc1}.

Initially, the strongest argument was based on how the data-acquisition system works~\cite{IceCube:2008qbc, Halzen:2010yj, IceCube:2016zyt}.  In short, the trigger for high-energy muons works by searching for coincidences of hits on eight nearby DOMs within 5 $\mu$s; if this occurs, all data within the whole detector within $\pm 10\ \mu$s is saved as one event.  Because real dimuons are simultaneous (and it takes 3 $\mu$s for a muon to travel 1 km), any real occurrence of a dimuon would be read out as one event.  However, we wondered if the dataset we use lists fitted tracks, not events.  We now know that this supposition is incorrect.

Also initially, another serious concern was that IceCube would be unable to fit two simultaneous tracks because the Cherenkov cones would be so similar.  Building on that point, for most of these events, the separation angles are comparable to the angular uncertainties.  However, it seemed reasonable that the {\it relative} angular uncertainty to separate two muons could be less than the {\it absolute} angular uncertainties on their directions.   This point remains uncertain to us.

Also initially, the only specific possible explanation for the events that seemed reasonable was that they may be some sort of ghost tracks, perhaps due to afterpulsing of photomultipliers, a phenomenon in which a photomultiplier that has collected a large amount of light then triggers again after a delay~\cite{IceCube:2010dpc, DEAP:2017fgw}.  Afterpulsing is a small effect ($\simeq 0.06$ of the original signal strength), with most of it occurring within a few $\mu$s, which, by the arguments in the paragraph two above, would cause it to be registered as part of the original muon event.  A smaller fraction of afterpulsing occurs at times around 11 $\mu$s, which could lead to a pattern of DOM hits that is registered as a separate event.  However, the distribution of hits would likely not form a good fit to a muon track, because the spread in time values would be large compared to that of a real muon, as the spread in afterpulsing times is large compared to the time for the muon to travel between DOMs (a small fraction of 1 $\mu$s).  In addition, the fake muon should be much fainter ($\sim 0.01$ less energetic for late hits) than the parent muon, which we do not find.  We now know that our counterarguments were correct.

Other possibilities were raised, though were unlikely.  First, dineutrino events from one cosmic-ray induced air shower. However, the event rate is only about one per 14 years~\cite{vanderDrift:2013zga}.  In any case, by the argument above, these would register as one event, due to being simultaneous.  Second, a track splitting into two, but this is rare, especially if the muons have comparable energies (and, again, these would be simultaneous).  Third, some kind of new-physics signal, e.g., double staus~\cite{Albuquerque:2003mi, Albuquerque:2006am, Ando:2007ds, ICECUBE:2013jjy, Kopper:2015rrp, Kopper:2016hhb, Meighen-Berger:2020eun, IC_stau_2021}.  

In conclusion, our 19 candidate events are not real dimuons, as shown by direct inspection of the events by IceCube. Even though we now know that these dimuon candidates are just misreconstructed muons, it remains an interesting mystery why the observed muon distributions were so close to our predictions. Only IceCube can shed light on this, and we encourage them to do so.  It might simply be a coincidence, but there might be issues that help understand backgrounds for future searches. Separately, we strongly encourage IceCube to develop a fitting algorithm specifically designed to find two simultaneous, nearby tracks.


\section{Discussions of Physical Potential}
\label{sec_dis}

In this section, we discuss more about the physics potential of dimuons.  In Sec.~\ref{sec_dis_QCD}, we show that dimuons could be important for measuring the strange-quark PDF, especially with current IceCube data.  In Sec.~\ref{sec_dis_WBP}, we show that the first detection of WBP could occur in the first few years of IceCube-Gen2 by using showerless starting dimuons.  In Sec.~\ref{sec_dis_other}, we show that dimuons are useful for neutrino-energy measurements and for characterizing backgrounds for new physics. 


\subsection{Implications for QCD}
\label{sec_dis_QCD}

\begin{figure*}[t]
\includegraphics[width=0.4\textwidth]{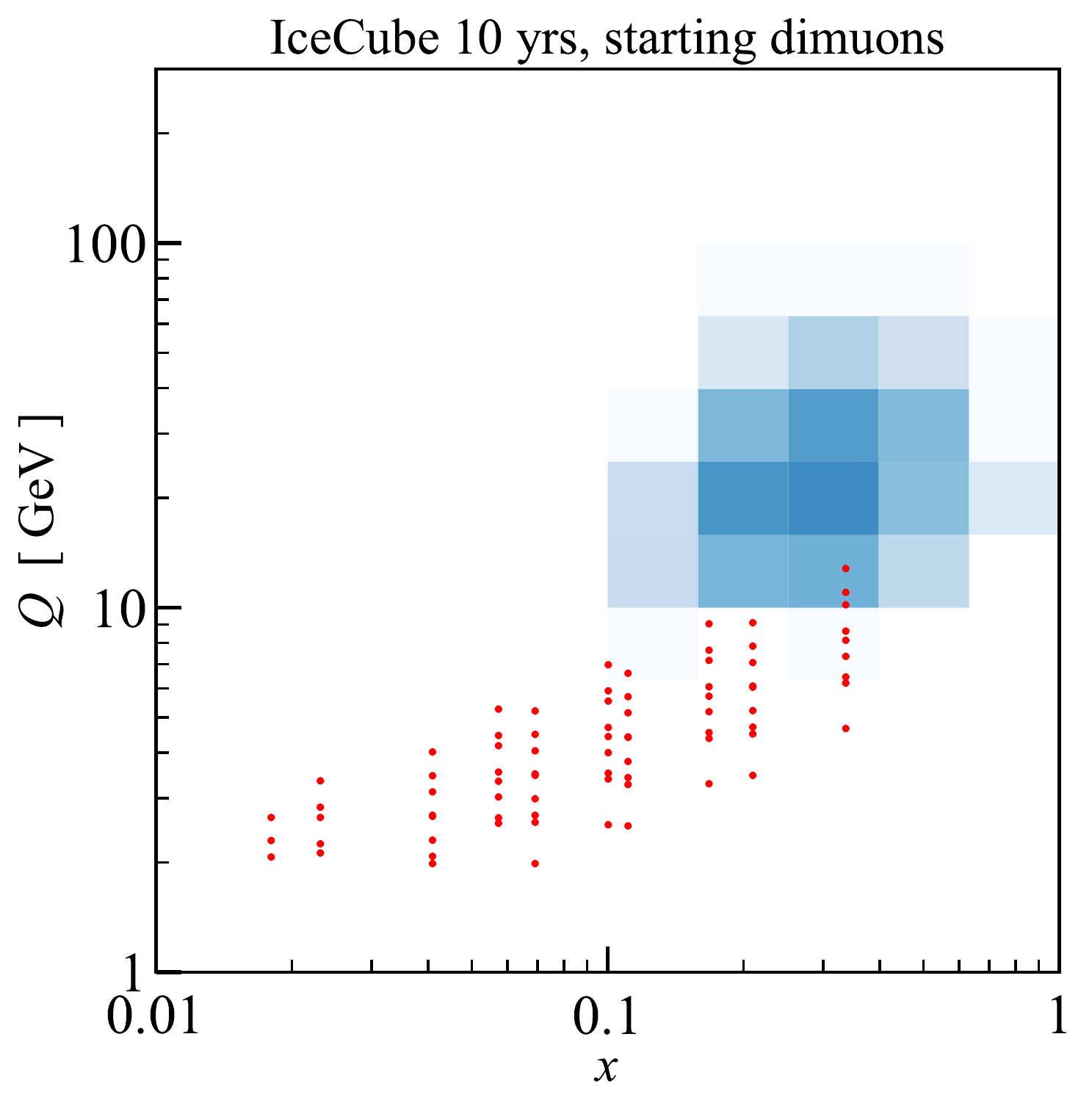}
\includegraphics[width=0.4\textwidth]{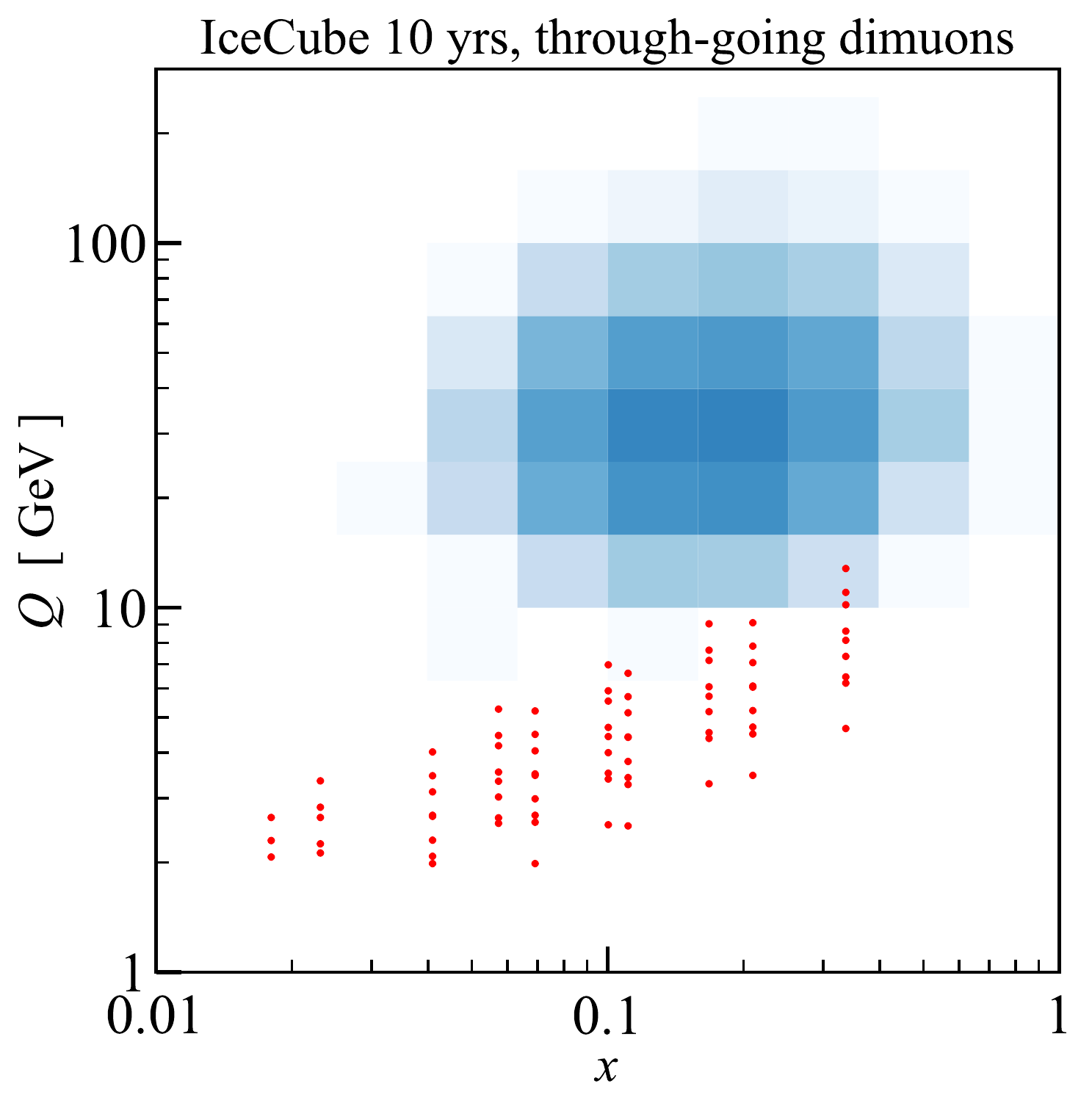}
\includegraphics[width=0.4\textwidth]{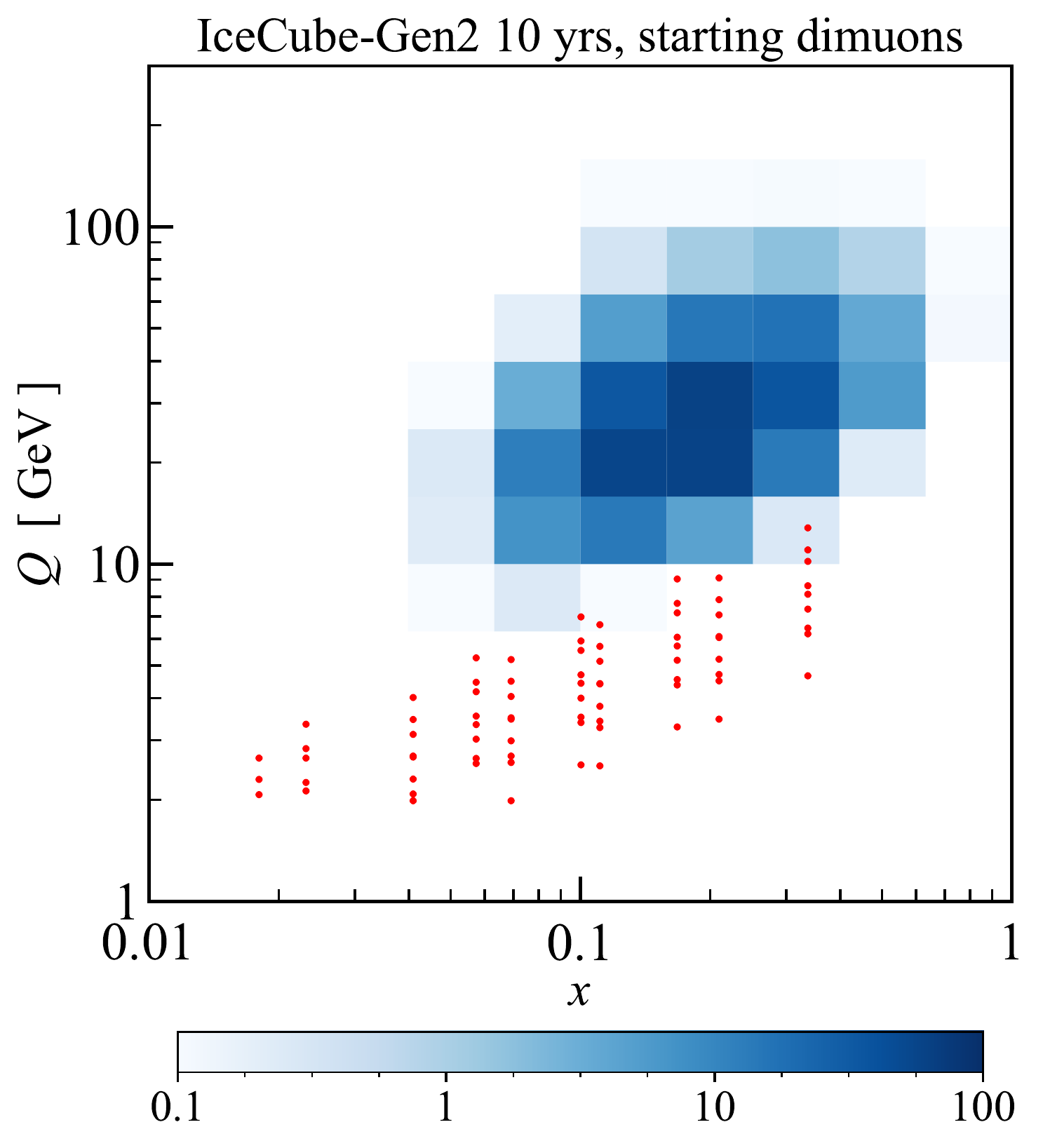}
\includegraphics[width=0.4\textwidth]{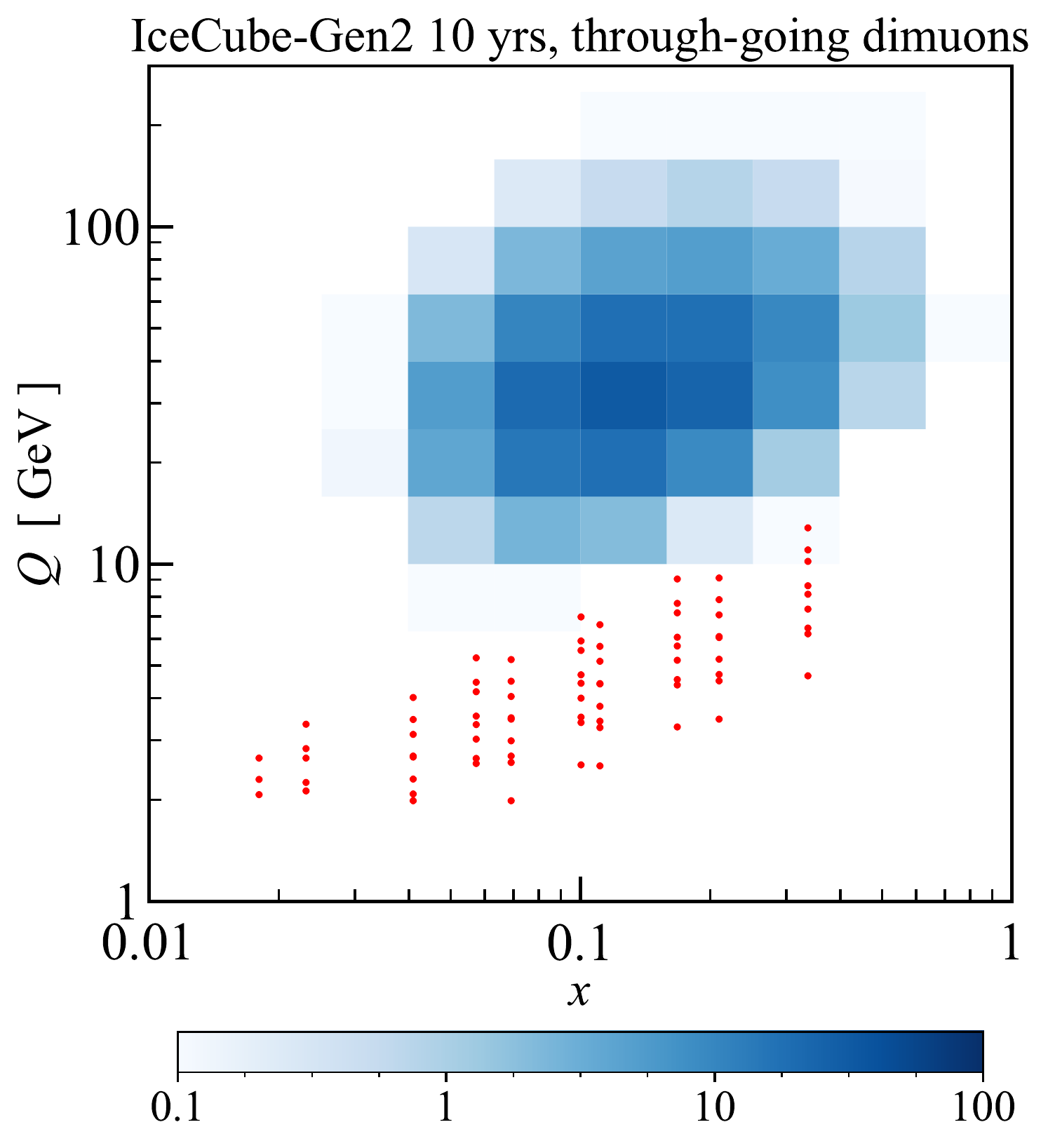}
\caption{
Distributions of parton-momentum fraction $x$ and factorization scale $Q$ for dimuon events.  Those in {\bf blue} are for our predicted dimuons (numbers of events in $x$-$Q$ bins) as titled in each panel, corresponding to the dimuons in Figs.~\ref{fig_dNdE_IceCube} and \ref{fig_dNdE_Gen2}.  All panels use the same color scale.  The dots are the data points of CCFR dimuons used for CT18 PDF set~\cite{CCFR:1994ikl, NuTeV:2001dfo, Hou:2019efy}.
}
\label{Fig_QCD}
\end{figure*}

\begin{figure*}[t!!]
\includegraphics[width=0.4\textwidth]{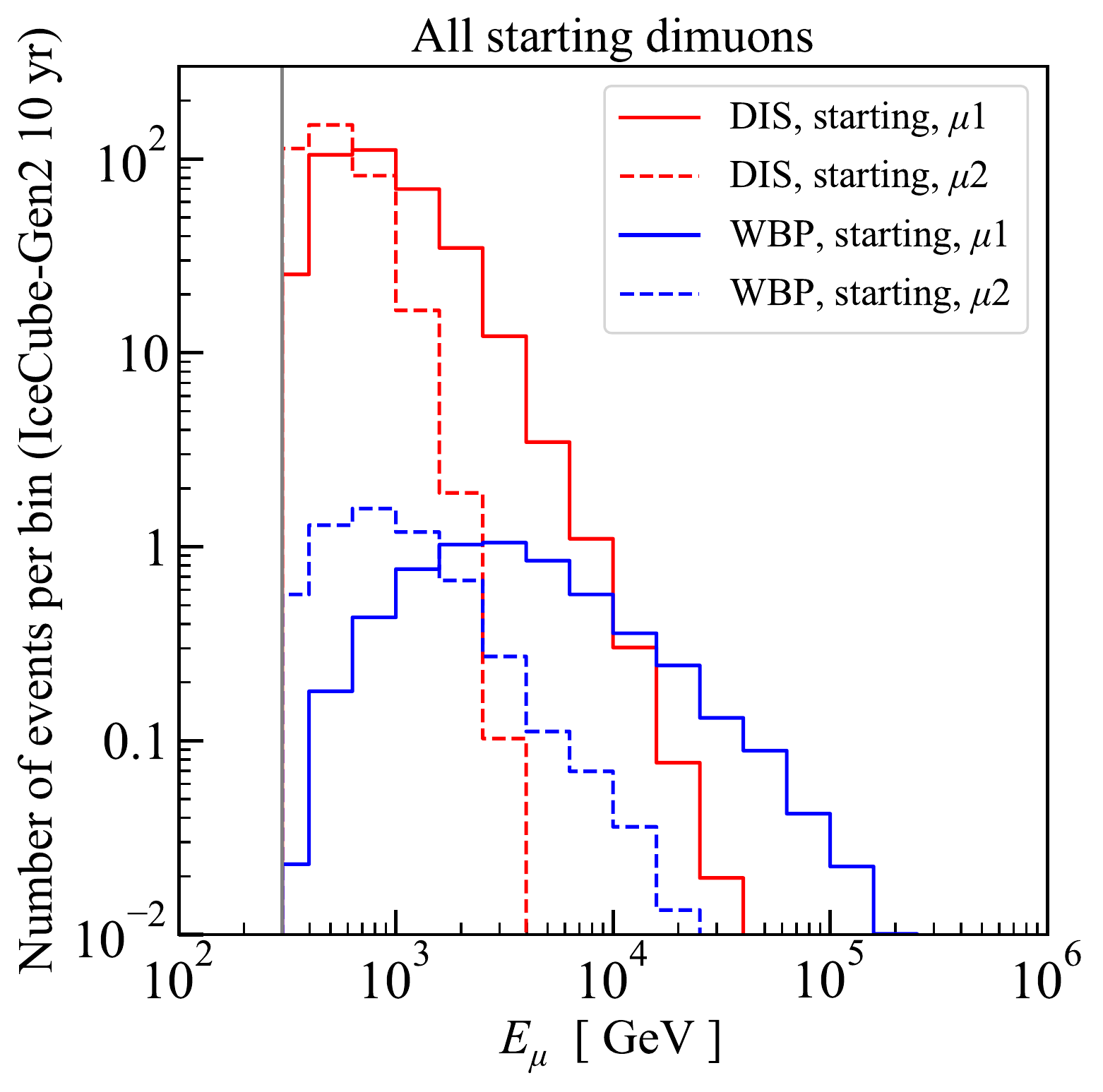}
\includegraphics[width=0.4\textwidth]{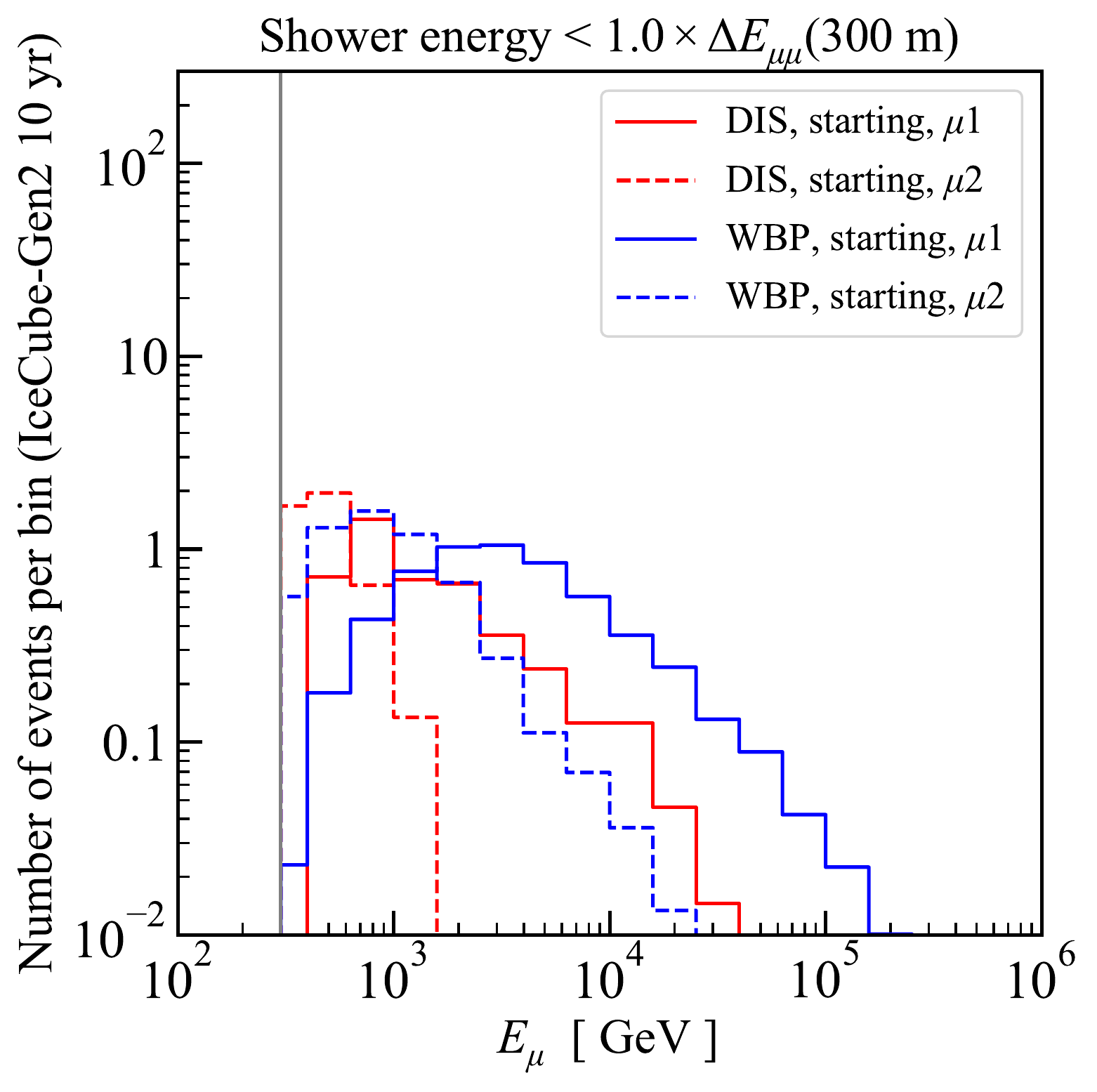}
\includegraphics[width=0.4\textwidth]{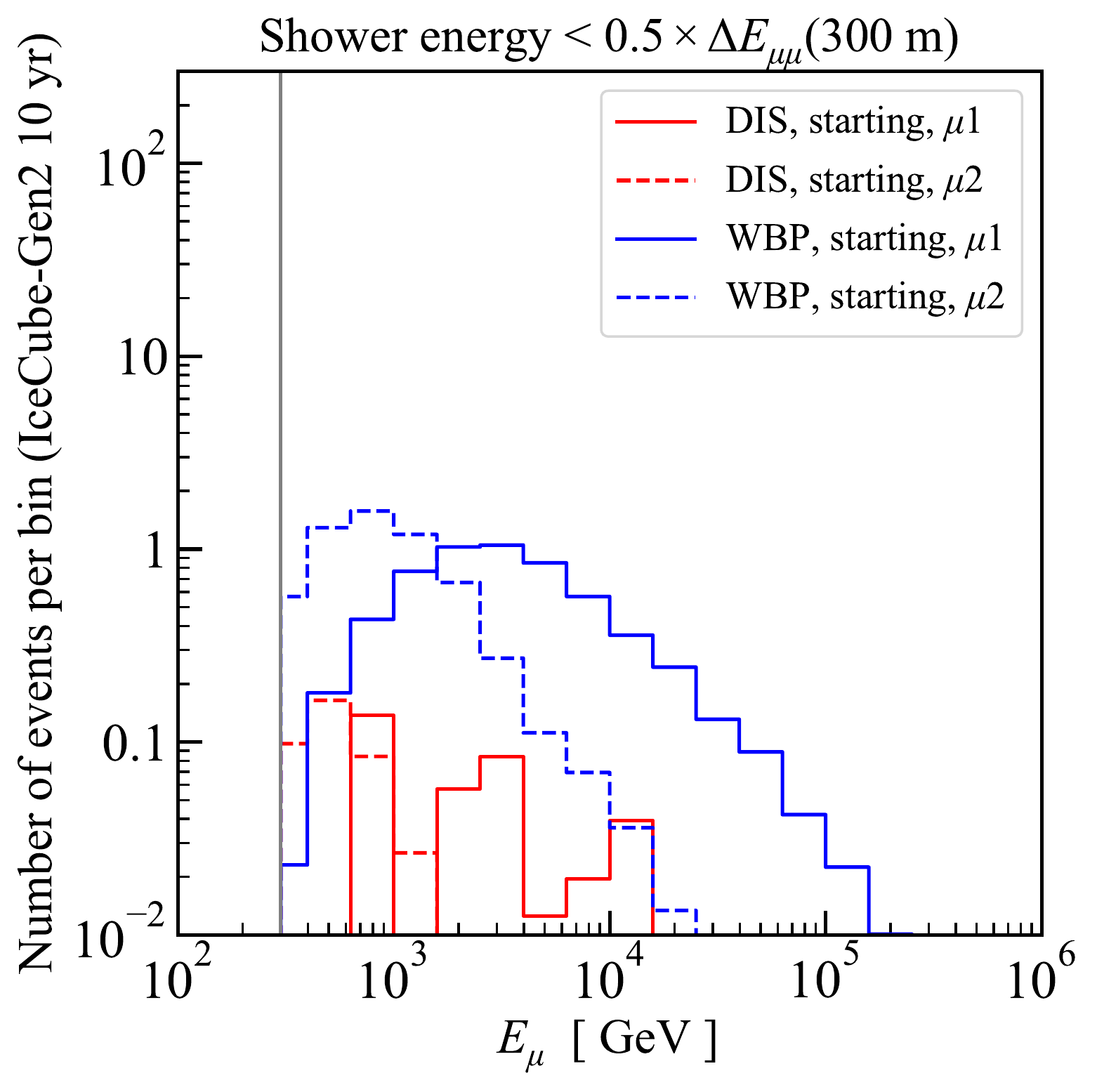}
\includegraphics[width=0.4\textwidth]{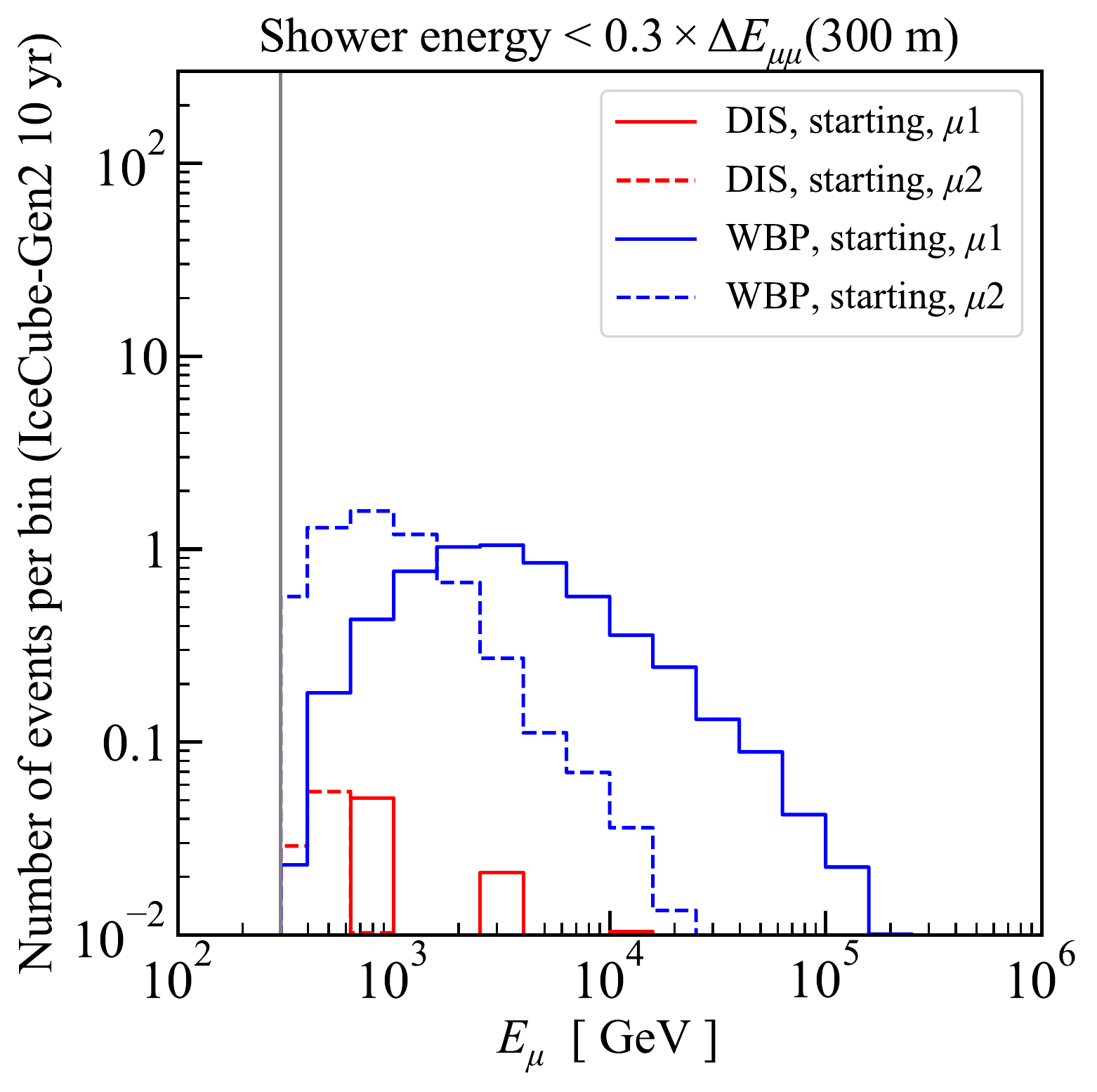}
\caption{
Starting dimuons for 10 years of IceCube-Gen2 with their shower energies below different thresholds, as noted in each panel.   $\Delta E_{\mu\mu}(300\ \rm m)$ is the average energy lost by the two muons in their initial path length of 300 m.  The {\bf top left panel} has a threshold of infinity and is the same as the left panel of Fig.~\ref{fig_dNdE_Gen2}.  The total number of starting WBP dimuons (signal) is about 5.8, and they are all showerless.  The total numbers of starting DIS dimuons (background) are about 370, 4.4, 0.4, and 0.1 for the top left, top right, bottom left, and bottom right panels, respectively. 
}
\label{fig_dNdE_Gen2_showerless}
\end{figure*}

Dimuon data from accelerator neutrino experiments have been shown to be very important for measuring the strange-quark PDF~\cite{DeLellis:2004ovi, Hou:2019efy, Faura:2020oom}.  Here we show that neutrino telescopes can do so at much higher factorization scales, by measuring dimuon cross sections at unprecedented high neutrino energies.  The numbers of events are less than those obtained in laboratory experiments, but the energy reach of neutrino telescopes is much greater.

We calculate the $x$-$Q$ distribution for the dimuons in IceCube (Fig.~\ref{fig_dNdE_IceCube}) and IceCube-Gen2 (Fig.~\ref{fig_dNdE_Gen2}), and compare with accelerator neutrino experiments.   Here $x$ is the parton-momentum fraction and $Q$ is the factorization scale.  For starting dimuons, this is
\begin{equation}
\begin{aligned}
\frac{d^2 N^{\rm st}_{\mu\mu}}{dx dQ} 
&=
T N_t 
\int_{}^{} d E_\nu
\frac{dF_\nu}{dE_\nu}(E_\nu) 
\frac{d^2 \sigma_{\mu\mu}^\text{cuts}}{dx dQ} (x, Q, E_\nu|E_{\mu2}'>E_\text{th})\,.
\end{aligned}
\end{equation}
For throughgoing muons (Appendix~\ref{app_thrgoCalc}), this is
\begin{equation}
\begin{aligned}
\frac{d^2 N_{\mu\mu}^\text{thr}}{dx dQ}
& = \frac{A_{\rm det} T N_A}{\beta} 
\int dE_\nu \frac{dF}{dE_\nu}(E_\nu)  \\
& \ \ \ \ \ \ \ 
\int_{E_{\rm th}}^{E_\nu} dE_{\mu2}'   \frac{d^3\sigma_{\mu\mu}^{\rm cuts}}{dx dQ dE_{\mu2}'} 
\ln \left( \frac{\alpha + \beta E_{\mu2}'}{\alpha + \beta E_{\rm th}} \right)  \,.
\end{aligned}
\end{equation}

Figure~\ref{Fig_QCD} shows the $x$-$Q$ distributions of dimuons in IceCube and IceCube-Gen2 from our calculation and those from the events from CCFR~\cite{CCFR:1994ikl, NuTeV:2001dfo, Hou:2019efy}, an accelerator neutrino experiment.   For simplicity, we do not show data from other accelerator experiments~\cite{NuTeV:2001dfo, Mason:2006qa, NOMAD:2013hbk, Hou:2019efy}, which cover a similar $x$-$Q$ range to CCFR.   Overall, IceCube and IceCube-Gen2 dimuons have $Q$ larger by one order of magnitude, because $Q \sim E_\nu$.  IceCube-Gen2 covers a larger region than IceCube because of its larger detector volume.  Neutrino telescopes will provide special opportunities to measure the strange-quark PDF on oxygen, whereas measurements with accelerator neutrino experiments have been on heavy targets (e.g., iron, lead, etc.)~\cite{Abramowicz:1982zr, CCFR:1994ikl, CHARMII:1998njb, NOMAD:2000sdx, NuTeV:2001dfo, CHORUS:2008vjb, NOMAD:2013hbk, Hou:2019efy}.  They are also complementary to LHC~\cite{ATLAS:2014jkm, CMS:2018dxg} (proton-proton collision) and HERA~\cite{ZEUS:2019oro} (electron-proton collision) data, which cover a similar $x$-$Q$ range. Importantly, the neutrino data uniquely probe the nuclear effects on the strangeness at larger $Q$, especially on the $s$-$\bar{s}$ asymmetry. Finally, we note that inelasticity measurements are also useful to measure the strange-quark PDF~\cite{IceCube:2018pgc}.

The regions covered by IceCube and IceCube-Gen2 dimuons are limited by the following factors.   According to DIS kinematics, $Q^2 \simeq 4 E_\nu E_\mu \sin^2 ({\theta_{\mu\mu}}/{2})$ and $x \simeq {4 E_\nu E_\mu \sin^2({\theta_{\mu\mu}}/{2})}/{(2 m_{N} E_{h})}$, where $m_N$ is the nucleon mass and $E_h$ the final-state hadronic energy. Therefore, $Q_{\rm max} \sim \text{maximum } E_\nu$, which depends on the detector size. Moreover, $Q_{\rm min} \sim E_{\rm th} \theta_{\mu\mu}^{\rm min}$ and $x_{\rm min} \sim E_{\rm th} (\theta^{\rm min}_{\mu\mu})^2$, which is determined by both energy and angular thresholds.   Importantly, lowering the angular threshold (i.e., the vertical spacing between DOMs) only moderately would significantly decrease $x_{\rm min}$, which, in principle, could be as small as $10^{-5}$ in IceCube/IceCube-Gen2.  This would also help lower $Q_{\rm min}$.


\subsection{Enabling the First Detection of WBP}
\label{sec_dis_WBP}

WBP interactions have not yet been detected, despite being second to only DIS in importance for high-energy neutrino interactions.  A critical feature of WBP events is that the nuclear interaction is typically soft, leading to events without hadronic showers, which is rare for DIS~\cite{Zhou:2019vxt, Zhou:2019frk}.  This is also true for the dimuon subset of WBP events.

We conservatively estimate the minimum detectable size of a hadronic shower in a dimuon event as follows.  The shower energy should be larger than a fraction $f$ (related to energy resolution) of the average energy lost by the muons $\Delta E_{\mu\mu}$ in their initial path length $L$ (precision related to the detector spacing).  From $dE/dX = -\alpha - \beta E$, we find that
\begin{equation}
\begin{aligned}
f \times \Delta E_{\mu\mu}(L) = f \times (E_{\mu1}'+E_{\mu2}'+2\epsilon)(1-e^{-\beta L \rho_{\rm ice}}) \, .
\label{eq_2nd_shth}
\end{aligned}
\end{equation}
Typically, $L \ll 1/(\beta \rho_{\rm ice}) \simeq 3.6$ km, so $f \times \Delta E_{\mu\mu}(L) \propto f \times L$, and $f$ and $L$ are degenerate.  Therefore, we fix $L = 300$ m, chosen to be somewhat larger than the spacing between two adjacent strings in IceCube-Gen2 (240~m).

Figure~\ref{fig_dNdE_Gen2_showerless} shows our starting dimuon spectra in IceCube-Gen2 after removing events with shower energies higher than certain fractions of the initial dimuon energy loss.  Different panels are for different $f$, which is degenerate with $L$. The results show that a cut on the shower energy reduces the DIS spectra at all $E_\mu$ values.  This can be understood from Eq.~(\ref{eq_2nd_shth}), which is approximately proportional to energy when $E_{\mu1}' + E_{\mu2}' \sim E_\nu \gg 2\epsilon \sim 1$ TeV.  Importantly, as expected, decreasing $f$ significantly lowers the number of showerless starting DIS dimuons. When $f < 0.3$, almost all of the showerless starting dimuons are from WBP, i.e., background free.

Our choice of shower-energy threshold should be reasonable for IceCube-Gen2. The energy resolution of IceCube, which is comparable to IceCube-Gen2, is $\simeq 10\%$ (systematics dominated) at least above a few TeV~\cite{Aartsen:2013jdh}, which is much better than the $f=0.3$ needed. Moreover, the topology of a shower, which looks like a blob, is quite different from that of a dimuon track. This will also help to remove the DIS starting dimuons.  On the other hand, the predicted DIS dimuon rate above $E_\nu \sim 10$~TeV would be lower if, as mentioned in Sec.~\ref{sec_pred_calc}, the interactions between $D$ mesons and matter are included, while the predicted WBP dimuon rate is unaffected. Finally, to make our calculation more realistic, it will be necessary to require not just that the muons are created in a showerless interaction, but also that the muons do not shower too much in the initial length segment.

We thus expect showerless starting dimuons in IceCube-Gen2 to be a background-free way to identify WBP events.   As we predict about 5.8 events from WBP in 10 years of IceCube-Gen2, the first detection could be accomplished within a few years.  This will test the standard model and constrain new physics.


\subsection{Other Implications}
\label{sec_dis_other}

For the same parent neutrino energy $E_\nu$, the reconstructed energy of a dimuon ($E_{\mu1}+E_{\mu2}$), is closer to $E_\nu$ than that of a single-muon event ($E_\mu$).  The first reason is that the dimuon loses much less energy than single-muon events before entering the detector, as the range of the production point is limited by $E_{\mu2}$, which is usually $\ll E_{\mu1} \sim E_\mu$.   A quantitative comparison can be obtained by comparing the number or acceptance of throughgoing events to starting events.  For single-muon events, the ratio is $\gtrsim 20$ while for dimuon events, it is only $\simeq 2$ (Table~\ref{tab_dimu_num}).  So, the range of the dimuons is $\lesssim (2/20) \lesssim 1/10$ of that of single-muon events.   The second reason is that the $E_{\mu2}$ could be used to estimate the energy transferred to the hadronic side.   This could help measure the spectrum~\cite{Aartsen:2015knd, Stettner:2019tok, Aartsen:2020aqd, IceCube:2020wum}, testing neutrino mixing~\cite{IceCube:2014flw, IceCube:2017ivd}, cross sections~\cite{Bustamante:2017xuy, IceCube:2017roe, IceCube:2020rnc, Robertson:2021yfz}, and more.
 
Dimuons could also be backgrounds for new-physics searches.  An intriguing example is the double staus in the supersymmetric models, induced by neutrino interactions.  These would also leave throughgoing double tracks in the detector and have been searched for by the IceCube Collaboration~\cite{Albuquerque:2003mi, Albuquerque:2006am, Ando:2007ds, ICECUBE:2013jjy, Kopper:2015rrp, Kopper:2016hhb, Meighen-Berger:2020eun, IC_stau_2021}.


\section{Conclusions}
\label{sec_concl}

Neutrino observatories play a critical role in astrophysics and particle physics~\cite{Davis:1968cp, Kamiokande-II:1987idp, Bionta:1987qt, Super-Kamiokande:1998kpq, SNO:2002tuh, Aartsen:2013jdh}.  High-energy neutrinos, a vibrant field, have been providing unique information on  astrophysics~\cite{Aartsen:2013jdh, Aartsen:2015knd, IceCube:2018cha, IceCube:2018dnn, Aartsen:2019fau, Aartsen:2020aqd, IceCube:2020wum}, standard-model particle physics~\cite{Glashow:1960zz, Lee:1960qv, Lee:1961jj, Seckel:1997kk, Alikhanov:2015kla, IceCube:2017roe, Bustamante:2017xuy, Zhou:2019vxt, Zhou:2019frk, IceCube:2020rnc, IceCube:2021rpz, Robertson:2021yfz}, and new physics~\cite{Beacom:2006tt, Yuksel:2007ac, Murase:2012xs, Feldstein:2013kka, Esmaili:2013gha, Murase:2015gea, IceCube:2016dgk, DiBari:2016guw, IceCube:2018tkk, Arguelles:2019ouk, ANTARES:2020leh, Lipari:2001ds, Cornet:2001gy, Beacom:2002vi, Ng:2014pca, Ioka:2014kca, Shoemaker:2015qul, Bustamante:2016ciw, Denton:2018aml, Bustamante:2020mep, Creque-Sarbinowski:2020qhz, Esteban:2021tub, Albuquerque:2003mi, Albuquerque:2006am, Ando:2007ds, ICECUBE:2013jjy, Kopper:2015rrp, Kopper:2016hhb, Meighen-Berger:2020eun, IC_stau_2021, Coloma:2017ppo} thanks to their unprecedented high energies, long propagation distances, and intervening high column densities (through Earth). On the other hand, those physics studies are based on only a few event classes, including muon tracks, showers, and double bangs.  New event classes should be studied to take more advantage of the data. 

For the theoretical contribution of this paper, we study dimuons in IceCube and IceCube-Gen2, focusing on production through DIS and WBP, the two most important processes.  We develop a theoretical framework to calculate the event rate and spectra of the starting and throughgoing dimuons (Sec.~\ref{sec_pred_calc}).  Our calculation shows that a dimuon-optimized search for 10 years of IceCube data  can find $\simeq 37+85$ DIS dimuons (starting + throughgoing) and $\simeq 0.3+6.0$ WBP dimuons (Fig.~\ref{fig_dNdE_IceCube}).  For 10 years of IceCube-Gen2, there will be $\simeq 370+230$ DIS dimuons, and $\simeq 5.8+22$ WBP dimuons (Fig.~\ref{fig_dNdE_Gen2}).    The above numbers are summarized in Table~\ref{tab_dimu_num}. These dimuons are almost all from atmospheric neutrinos.  We estimate the backgrounds caused by two coincident single muons (Sec.~\ref{sec_calc_Ncoinc}), finding them negligible for all but southern-sky throughgoing dimuons, but which could be substantially reduced.

For the observational contribution of this paper, we analyze 10 years of IceCube's publicly available data~\cite{IceCube:2021xar, data_webpage} (Sec.~\ref{sec_data}).  Importantly, we find 19 candidate dimuon events.  The IceCube public dataset is optimized for point-source searches instead of dimuon searches and has strong cuts especially at lower energies, so we find fewer events than we predict. 
{\it While these events have been found to be caused by an internal reconstruction error, they are an important background, previously unknown, for dimuon and other searches.}

Dimuons have important physics potential.  They provide new information about QCD, especially the strange-quark PDF~\cite{DeLellis:2004ovi, Hou:2019efy, Faura:2020oom}, probing factorization scales 10 times higher than accelerator neutrino data.  For IceCube, this measurement can be done with current data.  In Sec.~\ref{sec_dis_WBP}, we show that showerless~\cite{Zhou:2019vxt, Zhou:2019frk} starting dimuons could lead to the first detection of WBP events within a few years of IceCube-Gen2, several decades after the WBP process was first calculated by Lee and Yang~\cite{Lee:1960qv, Lee:1961jj}.  Finally, we note in Sec.~\ref{sec_dis_other} that the dimuons are better for reconstructing parent neutrino energies than single-muon events, and they are also a background for some new-physics scenarios.

The ideas and calculations in this paper can be applied to other neutrino telescopes, including KM3NeT~\cite{Adrian-Martinez:2016fdl}, Baikal-GVD~\cite{Baikal-GVD:2018isr}, P-ONE~\cite{P-ONE:2020ljt}, and especially IceCube-Gen2~\cite{IceCube-Gen2:2020qha}.  The physics characteristics of the dimuon events could even shape choices about detector design.  Lower-energy detectors are also worth studying, such as Super-Kamionkande~\cite{SK_web}, DUNE~\cite{DUNE:2015lol}, DeepCore~\cite{IceCube:2011ucd}, and the IceCube Upgrade~\cite{IceCube-PINGU:2014okk}.  Their lower energy and angular thresholds could complement their smaller volumes.  In these detectors, dimuons from tridents are especially important~\cite{Altmannshofer:2014pba, Magill:2016hgc, Ge:2017poy, Ballett:2018uuc, Gauld:2019pgt, Altmannshofer:2019zhy, Zhou:2019vxt, Zhou:2019frk}.  Moreover, the much lower angular threshold could be important to probe small-$x$ regions for the strange-quark PDF (see Sec.~\ref{sec_dis_QCD}).  In addition, new theoretical studies are needed to develop techniques to isolate dimuon events, to exploit their physics potential, and to explore new-physics scenarios for which dimuons are backgrounds.

The continued success of neutrino physics and astrophysics depends on developing new tools to get the most out of the data.  An important part of that is developing new event classes, although it is not easy (e.g., compare the decades of work on tau double-bang events).


\section*{Acknowledgments}

We are grateful for helpful discussions with Bob Bernstein, Tom Gaisser, Alfonso Garcia Soto, Xiaoyuan Huang, Claudio Kopper, Stephen Mrenna, Aditya Parikh, Carsten Rott, and especially Francis Halzen, Spencer Klein, Keping Xie, and the many members of the IceCube Collaboration who helped to quickly investigate our candidate events.  We used {\tt FeynCalc}~\cite{Mertig:1990an, Shtabovenko:2016sxi}, {\tt MadGraph}~\cite{Alwall:2014hca}, and {\tt Pythia}~\cite{Sjostrand:2014zea} for some calculations.  B.Z. was supported by the Simons Foundation.  J.F.B. was supported by National Science Foundation Grant No. PHY-2012955.


\onecolumngrid

\appendix

\newpage

\section{Dimuon Candidates in IceCube Data}
\label{app_dimu}

\begin{figure*}[t!]
\includegraphics[width=0.4\columnwidth]{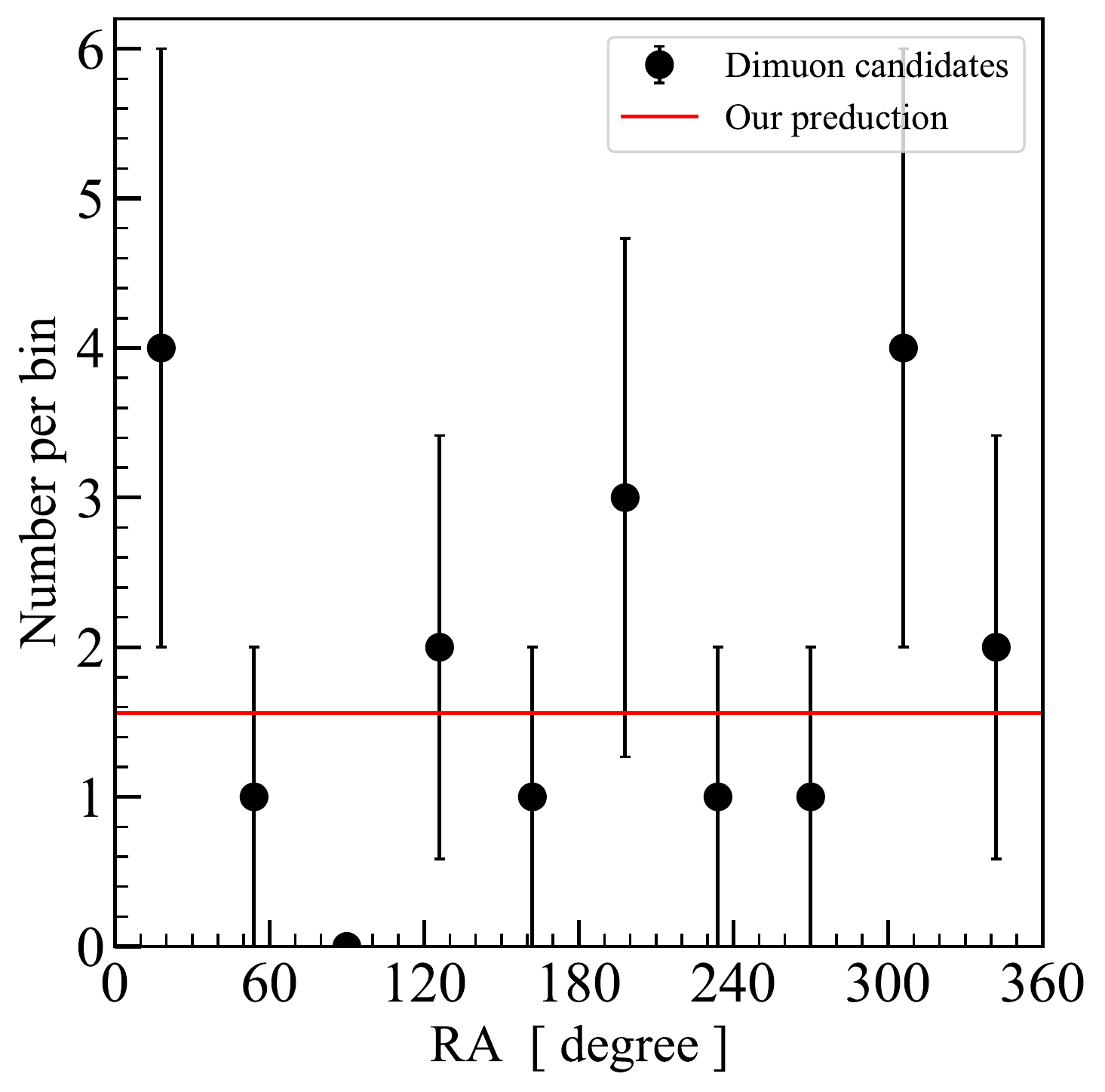}
\includegraphics[width=0.4\columnwidth]{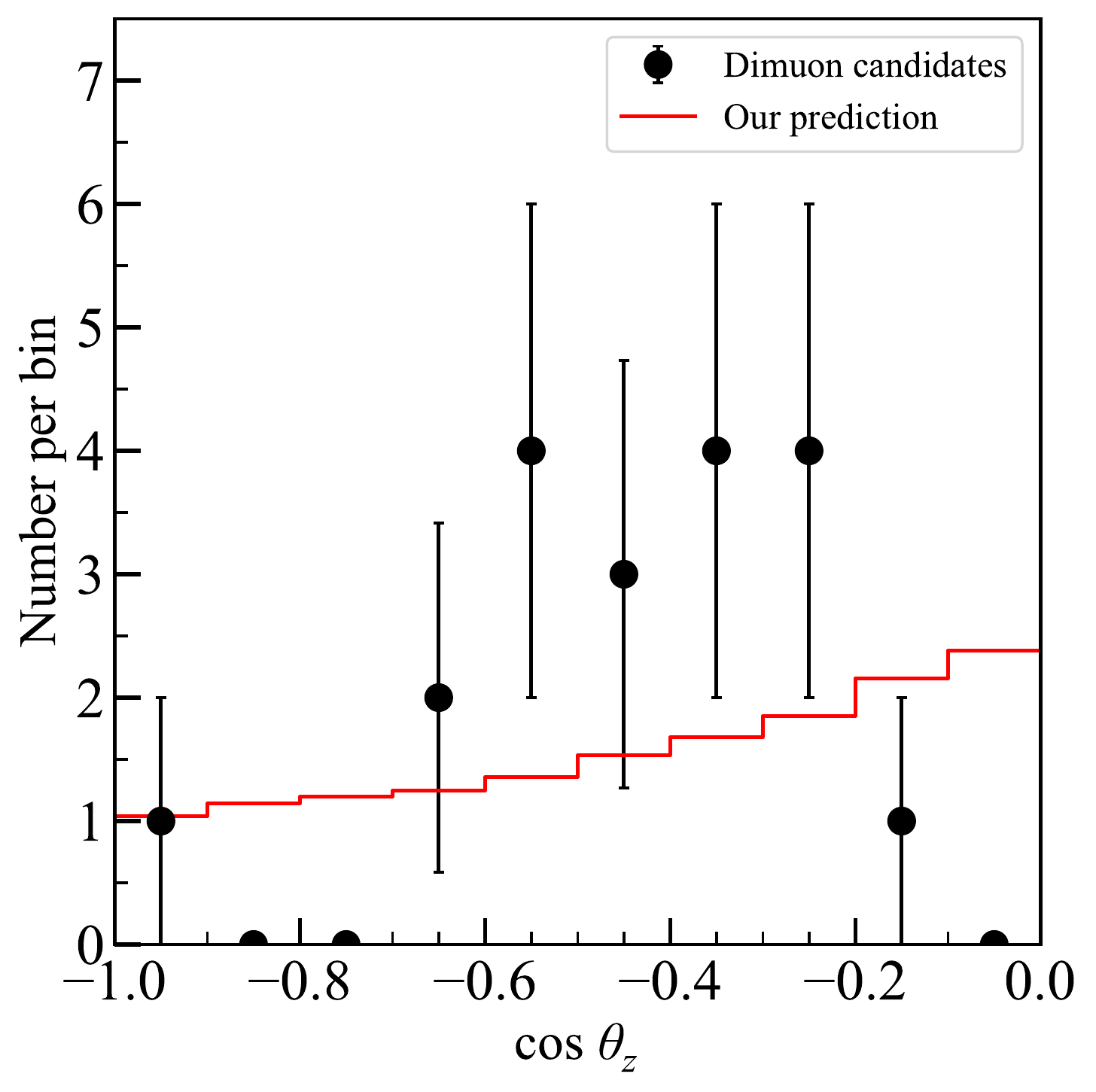}
\caption{
Distributions in RA and $\cos \theta_z$ of the 19 dimuon candidates, along with our predictions, where
$\rm RA = (RA1+RA2)/2$ and $ \cos \theta_z = \cos([\theta_{z,1} + \theta_{z,2}]/2)$, where $\theta_{z, 1/2} \simeq \rm Dec1/2 + 90^\circ$ for IceCube's location.  The data agree well with predictions.
}
\label{fig_dimu_RADec}
\end{figure*}

In this appendix, we give more details about the 19 dimuon candidates that we find in the IceCube 2008--2018 data~\cite{IceCube:2021xar, data_webpage}.  Table~\ref{tab_data_dimu_list} lists the details of these events. Figure~\ref{fig_dimu_RADec} shows the sky angular distributions of the dimuon candidates compared to our predictions, finding good agreement.

\begin{table*}[h!]
\caption{
Each row is one event (two muons).  Column names ending with ``1'' and ``2'' are for $\mu1$ and $\mu2$, respectively.  MJD1/2 are their arrival times in the unit of modified Julian day (MJD).  $E_{\mu1/2}$ are their energies (energy proxies).  RA1/2 and Dec1/2 are their arrival directions in the equatorial coordinate system (for IceCube's location, $\rm zenith\ angle = Dec + 90^\circ$).  AngErr1/2 are their $1\sigma$ angular errors.  AngDis is the angular separation between each $\mu1$ and $\mu2$, with uncertainty of $\rm DisErr = \sqrt{AngErr1^2+AngErr2^2}$.  We quote the data as obtained from the IceCube, noting that most values are given with more digits than warranted by the uncertainties.
}
\label{tab_data_dimu_list}
\medskip
\renewcommand{\arraystretch}{1.1} 
\centering
\begin{tabular*}{0.955\textwidth}{c|c|c|c|c|c|c|c|c|c|c|c}
    MJD1 [day] & MJD2 (= MJD1) & $E_{\mu1}$ [TeV] & $E_{\mu2}$ & RA1 [deg] & RA2 & Dec1 & Dec2 & AngErr1 & AngErr2 & AngDis & DisErr  \\	\hline \hline

56068.26557772	&	56068.26557772	&	1.23	&	1.05	&	25.065	&	25.860	&	18.168	&	18.466	&	0.38	&	1.85	&	0.81	&	1.89		 \\ \hline
56115.78056499	&	56115.78056499	&	2.29	&	0.65	&	296.835	&	296.891	&	41.777	&	46.922	&	3.10	&	0.41	&	5.15	&	3.13	 \\ \hline
56235.14756523	&	56235.14756523	&	2.19	&	2.19	&	179.781	&	185.182	&	20.271	&	28.274	&	2.50	&	1.57	&	9.39	&	2.95	 \\ \hline
56582.68675378	&	56582.68675378	&	2.29	&	1.35	&	120.687	&	121.892	&	26.630	&	24.994	&	1.47	&	0.78	&	1.96	&	1.66		 \\ \hline
56653.19502448	&	56653.19502448	&	3.31	&	1.48	&	48.106	&	47.781	&	30.840	&	30.100	&	0.75	&	1.19	&	0.79	&	1.41		 \\ \hline
56784.87114671	&	56784.87114671	&	1.35	&	0.35	&	126.690	&	126.357	&	69.524	&	70.871	&	1.97	&	2.83	&	1.35	&	3.45	 \\ \hline
56813.78701082	&	56813.78701082	&	0.91	&	0.83	&	184.136	&	181.708	&	31.627	&	31.957	&	3.01	&	0.83	&	2.09	&	3.12		 \\ \hline
56895.78341718	&	56895.78341718	&	1.91	&	0.79	&	295.288	&	303.817	&	14.387	&	16.670	&	1.94	&	1.61	&	8.53	&	2.52	 \\ \hline
56932.15214130	&	56932.15214130	&	1.70	&	0.98	&	175.546	&	173.549	&	36.710	&	35.972	&	1.17	&	0.86	&	1.77	&	1.45		 \\ \hline
56940.02405671	&	56940.02405671	&	5.13	&	3.72	&	1.404	&	0.541	&	11.716	&	9.353	&	3.13	&	2.38	&	2.51	&	3.93		 \\ \hline
57214.99298310	&	57214.99298310	&	1.51	&	0.83	&	13.089	&	14.760	&	39.101	&	39.034	&	3.50	&	0.85	&	1.30	&	3.60		 \\ \hline
57376.46221142	&	57376.46221142	&	1.66	&	1.55	&	326.795	&	328.022	&	17.543	&	15.199	&	2.11	&	1.15	&	2.62	&	2.40		 \\ \hline
57461.19606500	&	57461.19606500	&	1.35	&	1.10	&	308.771	&	307.274	&	31.268	&	30.077	&	1.08	&	1.37	&	1.75	&	1.74	 \\ \hline
57499.81363094	&	57499.81363094	&	5.89	&	1.70	&	199.430	&	201.527	&	16.454	&	15.029	&	2.55	&	1.30	&	2.47	&	2.86	 \\ \hline
57560.74070687	&	57560.74070687	&	1.74	&	0.79	&	219.566	&	219.023	&	12.582	&	13.008	&	1.62	&	0.74	&	0.68	&	1.78		 \\ \hline
57650.26270928	&	57650.26270928	&	6.17	&	2.40	&	256.189	&	255.088	&	19.588	&	20.293	&	2.03	&	0.77	&	1.25	&	2.17	 \\ \hline
57661.79317519	&	57661.79317519	&	1.45	&	0.91	&	24.276	&	21.095	&	23.145	&	24.317	&	1.72	&	2.22	&	3.14	&	2.81		 \\ \hline
58003.09416087	&	58003.09416087	&	2.29	&	1.23	&	349.095	&	345.586	&	21.328	&	19.554	&	2.17	&	1.30	&	3.74	&	2.53		 \\ \hline
58266.46093610	&	58266.46093610	&	2.63	&	1.48	&	296.881	&	294.994	&	19.596	&	20.896	&	1.57	&	1.45	&	2.20	&	2.14		 \\ \hline
\end{tabular*}
\end{table*}

\newpage

\section{Calculational Framework for Throughgoing Dimuons}
\label{app_thrgoCalc}

In this appendix, we develop the first calculational framework to calculate the event rate and spectra of throughgoing dimuons.  This framework can also be adapted for other particles and materials (simply change $\alpha$ and $\beta$), such as distaus from supersymmetric models~\cite{Albuquerque:2003mi, Albuquerque:2006am, Ando:2007ds, ICECUBE:2013jjy, Kopper:2015rrp, Kopper:2016hhb, Meighen-Berger:2020eun, IC_stau_2021}.  

The setup is as follows. A neutrino interacts outside the detector and produces two muons, with initial energies $E_{\mu1}'$ and $E_{\mu2}'$.  We denote the more energetic muon as $\mu1$ and the less energetic muon as $\mu2$.  The muons travel while losing energy in matter, eventually entering the detector with energies $E_{\mu1}$ and $E_{\mu2}$.  Because the detector has energy and angular thresholds, only those muons with enough energy (and large enough angular separation) can be identified as dimuons.

The number of detectable throughgoing dimuons, $N_{\mu\mu}^\text{thr}$, is limited by how many $\mu2$ can reach the detector above the energy threshold ($E_\text{th}$), i.e., 
\begin{equation}
\begin{aligned}
N_{\mu\mu}^\text{thr} 
& = A_\text{det} T
\int_{E_\text{th}}^{\infty} d E_{\mu2}
\int_{E_{\mu2}}^{\infty} dE_\nu \frac{d F_\nu}{dE_\nu}(E_\nu) \frac{dP_{\rm tot}}{dE_{\mu2}}(E_{\mu2},E_\nu) \\
& = A_\text{det} T
\int_{E_\text{th}}^{\infty} d E_{\mu2}
\int_{E_{\mu2}}^{\infty} dE_\nu \frac{d F_\nu}{dE_\nu}(E_\nu)  
\int_{E_{\mu2}}^{E_\nu} d E_{\mu2}'
\int_{0}^{\infty} dX
N_A  \frac{d \sigma^\text{cuts}_{\mu\mu}}{dE_{\mu2}'}(E_{\mu2}', E_\nu) \,  \frac{d P_2(E_{\mu2}, E_{\mu2}', X)}{dE_{\mu2}} ,
\label{eq_Nmm_4ints}
\end{aligned}
\end{equation}
where $A_\text{det}$ is the cross-sectional area of the detector (as a good approximation, we assume it to be direction-independent), $T$ the observation time, ${d F_\nu}/{dE_\nu}$ is the neutrino flux in a certain energy and zenith range after considering absorption by the Earth, and $dP_{\rm tot}/dE_{\mu2}$ is the probability density for a neutrino with energy $E_\nu$ to produce a $\mu2$ with energy $E_{\mu2}$ entering the detector.  In the second line, $X$ (in column-density units of $\rm g/cm^2$) is the path length of a muon between its production point and where it enters the detector, $N_A = 6.02 \times 10^{-23} \, \rm g^{-1}$ is the Avogadro number, and $\sigma^\text{cuts}_{\mu\mu}$ is the dimuon cross section after angular-separation cuts, and $dP_2/dE_{\mu2}$ is the probability density (specified below) for a $\mu2$ with initial energy $E_{\mu2}'$ to reach the detector with energy $E_{\mu2}$ after traveling a distance $X$.  From Eq.~(\ref{eq_Nmm_4ints}), we can obtain the spectrum of $E_{\mu2}$, i.e., $d N_{\mu\mu}^\text{thr} / d E_{\mu2}$. 

Then, for the spectrum of $\mu1$, we can start with a similar formula, requiring that both $\mu1$ and its associated $\mu2$ enter the detector with energies above $E_\text{th}$, 
\begin{equation}
\begin{aligned}
\frac{d N_{\mu\mu}^\text{thr} }{d E_{\mu1}}
& = A_\text{det} T
\int_{E_{\mu1}}^{\infty} dE_\nu \frac{d F_\nu}{dE_\nu}(E_\nu) 
\frac{dP_{\rm tot}}{dE_{\mu1}}(E_{\mu1}, E_{\mu2} > E_\text{th}, E_\nu)  \\
& = A_\text{det} T
\int_{E_{\mu1}}^{\infty} dE_\nu \frac{d F_\nu}{dE_\nu}(E_\nu) 
\int_{E_{\mu1}}^{E_\nu} d E_{\mu1}'
\int_{E_\text{th}}^{E_{\mu1}'} d E_{\mu2}
\int_{E_{\mu2}}^{E_{\mu1}'} d E_{\mu2}'
\int_{0}^{\infty} d X  \\
& \ \ \ \ \ \ \ \ \ \ \ \ \ \ \ \ \ \ \ \ \ \ \ \ \ \ \ \ \ \ \ \ \ \ \ \ \ \ \ \ \ \ \ \ \ \ \ \ \ \ \ \  N_A \frac{d^2\sigma^\text{cuts}_{\mu\mu}}{d E_{\mu1}' d E_{\mu2}'}
\frac{d P_2}{d E_{\mu2}}(E_{\mu2}, E_{\mu2}', X) 
\frac{d P_1}{d E_{\mu1}}(E_{\mu1}, E_{\mu1}', X),
\label{eq_dNmmdEm1_5ints}
\end{aligned}
\end{equation}
where the notation is very similar to that above.

To simplify the integrals above, we need to specify $d P_{1}/d E_{\mu1}$ and $d P_{2}/d E_{\mu2}$.  Here we approximate them, assuming that the muon energy loss takes its average value.  (In principle, these expressions can be obtained from simulations, taking into account stochastic fluctuations.)  To a very good approximation~\cite{Dutta:2000hh, Groom:2001kq, ParticleDataGroup:2020ssz}, 
\begin{equation}
\begin{aligned}
\frac{d P}{d E_{\mu}}(E_{\mu}, E_{\mu}', X)  
= \delta(E_{\mu} - \langle E_{\mu}(X) \rangle) 
= \delta(E_{\mu}- [ (E_{\mu}'+\epsilon) e^{-\beta X} - \epsilon ]),
\end{aligned}
\end{equation}
where $\langle E_{\mu}(X) \rangle$ denotes the average energy of the muon (with initial energy $E_\mu'$) after losing energy over a distance $X$ and $\epsilon = \alpha/\beta$ is the critical energy of the muon (above which the radiative energy loss dominates).  Then, for example, $d P_2 / d E_{\mu2}$ can be written as
\begin{equation}
\begin{aligned}
\frac{d P_2}{d E_{\mu2}}(E_{\mu2}, E_{\mu2}', X)  = \delta(E_{\mu2}-\langle E_{\mu2}(X) \rangle) 
= 
\frac{\delta(X-X_0)}{\alpha+\beta E_{\mu2}} 
, \text{ with } X_0 = \frac{1}{\beta} \ln \left( \frac{E_{\mu2}' + \epsilon}{E_{\mu2} + \epsilon} \right) \,.
\label{eq_dP2dEmu2}
\end{aligned}
\end{equation}

Returning to Eq.~(\ref{eq_Nmm_4ints}), this now reduces to
\begin{equation}
\begin{aligned}
N_{\mu\mu}^\text{thr} 
& = A_\text{det} T N_A
\int_{E_\text{th}}^{\infty} d E_{\mu2}
\frac{1}{\alpha+\beta E_{\mu2}} 
\int_{E_{\mu2}}^{\infty} dE_\nu \frac{d F_\nu}{dE_\nu}(E_\nu)  
\int_{E_{\mu2}}^{E_\nu} d E_{\mu2}'
\frac{d \sigma^\text{cuts}_{\mu\mu}}{dE_{\mu2}'}(E_{\mu2}', E_\nu),
\label{eq_Nmm_3ints}
\end{aligned}
\end{equation}
from which we can also get $d N_{\mu\mu}^\text{thr} / d E_{\mu2}$.  After changing the order of integration, the integration over $E_{\mu2}$ can be evaluated analytically, which gives
\begin{equation}
\begin{aligned}
N_{\mu\mu}^\text{thr} 
& = \frac{A_\text{det} T N_A}{\beta}
\int_{E_\text{th}}^{\infty} d E_{\mu2}'
\int_{E_{\mu2}'}^{\infty} dE_\nu \frac{d F_\nu}{dE_\nu}(E_\nu)  
\frac{d \sigma^\text{cuts}_{\mu\mu}}{dE_{\mu2}'}(E_{\mu2}', E_\nu)
\ln{ \left( \frac{\alpha+\beta E_{\mu2}'}{\alpha+\beta E_\text{th}} \right)  } .
\label{eq_Nmm_2ints}
\end{aligned}
\end{equation}

For the spectrum of $\mu1$, Eq.~(\ref{eq_dNmmdEm1_5ints}) reduces in a similar way to
\begin{equation}
\begin{aligned}
\frac{d N_{\mu\mu}^\text{thr}}{d E_{\mu1}}
& = A_\text{det} T N_A
\int_{E_{\mu1}}^{\infty} dE_\nu \frac{d F_\nu}{dE_\nu}(E_\nu) 
\int_{E_{\mu1}}^{E_\nu} d E_{\mu1}'
\int_{E_\text{th}}^{E_{\mu1}'} d E_{\mu2}
\int_{E_{\mu2}}^{E_{\mu1}'} d E_{\mu2}' \,  \\
&\ \ \ \ \ \ \ \ \ \ \  \frac{d^2\sigma^\text{cuts}_{\mu\mu}}{d E_{\mu1}' d E_{\mu2}'} (E_{\mu1}', E_{\mu2}', E_\nu)
\frac{1}{\alpha+\beta E_{\mu2}} 
\frac{d P_1}{d E_{\mu1}}(E_{\mu1}, E_{\mu1}', X_0)  \, .
\label{eq_dNmmdEm1_4ints}
\end{aligned}
\end{equation}
Similar to Eq.~(\ref{eq_dP2dEmu2}), $dP_1/d E_{\mu1}$ can be written as
\begin{equation}
\begin{aligned}
\frac{d P_1}{d E_{\mu1}}(E_{\mu1}, E_{\mu1}', X_0) 
& =  \frac{E_{\mu2}'+\epsilon}{E_{\mu1}'+\epsilon} 
\delta \left( E_{\mu2} - \left[ (E_{\mu2}'+\epsilon) \frac{E_{\mu1}+\epsilon}{E_{\mu1}'+\epsilon} - \epsilon \right] \right)  .
\end{aligned}
\end{equation}
Note that the delta function is fixed at $X_0$ by  Eq.~(\ref{eq_dP2dEmu2}).  Plugging this into Eq.~(\ref{eq_dNmmdEm1_4ints}), we obtain
\begin{equation}
\begin{aligned}
\frac{d N_{\mu\mu}^\text{thr}}{d E_{\mu1}}
& = 
\frac{A_\text{det} T N_A}{\alpha+\beta E_{\mu1}}
\int_{E_{\mu1}}^{\infty} dE_\nu \frac{d F_\nu}{dE_\nu}
\int_{E_{\mu1}}^{E_\nu} d E_{\mu1}'   
\int_{E_{\mu2, \rm th}'}^{E_{\mu1}'} d E_{\mu2}' \, 
\frac{d^2\sigma^\text{cuts}_{\mu\mu}}{d E_{\mu1}' d E_{\mu2}'} (E_{\mu1}', E_{\mu2}', E_\nu)  \,,
\end{aligned}
\end{equation}
where $E_{\mu2}' > E_{\mu2, \text{th}}' = \frac{E_{\mu1}'+\epsilon}{E_{\mu1}+\epsilon}(E_\text{th}+\epsilon) - \epsilon$.
As a cross check, integrating over $E_{\mu1}$ and $E_{\mu1}'$ recovers Eq.~(\ref{eq_Nmm_2ints}).

Finally, the distributions in $x$ and $Q$ can be calculated as
\begin{equation}
\begin{aligned}
\frac{d^2 N_{\mu\mu}^\text{thr}}{dx dQ}
& = A_{\rm det} T \int dE_\nu \frac{dF}{dE_\nu}(E_\nu) \frac{d^2P_{\rm tot}}{dxdQ}(x, Q, E_\nu | E_{\mu2}>E_{\rm th}) \\
& = A_{\rm det} T \int dE_\nu \frac{dF}{dE_\nu}(E_\nu) 
\int_{E_{\rm th}}^{E_\nu} dE_{\mu2} \int_{E_{\mu2}}^{E_\nu} dE_{\mu2}' 
\int_0^{\infty} dX N_A \frac{d^3\sigma_{\mu\mu}^{\rm cuts}}{dx dQ dE_{\mu2}'} \frac{d P_2}{d E_{\mu2}}(E_{\mu2}, E_{\mu2}', X) \\
& = 
\frac{A_{\rm det} T N_A}{\beta}
\int dE_\nu \frac{dF}{dE_\nu}(E_\nu)  
\int_{E_{\rm th}}^{E_\nu} dE_{\mu2}'   
\ln\left(\frac{E_{\mu2}' + \epsilon}{E_{\rm th} + \epsilon}\right)
\frac{d^3 \sigma_{\mu\mu}^{\rm cuts} }{dx dQ dE_{\mu2}'} (x, Q, E_{\mu2}', E_\nu) \, ,
\label{eq_dNdxdQ}  
\end{aligned}
\end{equation}
and the distribution in $\theta_{\mu\mu}$ can be calculated as
\begin{equation}
\begin{aligned}
\frac{d N_{\mu\mu}^\text{thr}}{d\theta_{\mu\mu}}
& = \frac{A_{\rm eff} T N_A}{\beta}  \int dE_\nu \frac{dF}{dE_\nu}(E_\nu) 
\int_{E_{\rm th}}^{E_\nu} dE_{\mu2}'   
\ln \left( \frac{\alpha + \beta E_{\mu2}'}{\alpha + \beta E_{\rm th}} \right)
\frac{d^2\sigma_{\mu\mu}^{\rm cuts}}{d\theta_{\mu\mu} dE_{\mu2}'} 
\label{eq_app_dNdtheta}  
\end{aligned}
\end{equation}
As a cross check, integrating over $x$ and $Q$ in Eq.~(\ref{eq_dNdxdQ}) or $\theta_{\mu\mu}$ in Eq.~(\ref{eq_app_dNdtheta}) recovers Eq.~(\ref{eq_Nmm_2ints}).

\twocolumngrid


\bibliography{references}

\end{document}